\documentclass[aps,prd,10pt,twocolumn,superscriptaddress,bibnotes,longbibliography,preprintnumbers]{revtex4-1}

\pdfoutput=1
\usepackage{hyperref}
\usepackage{graphicx}
\usepackage{amsfonts,amsmath,amssymb,bm,bbm}
\usepackage{color}
\usepackage{enumitem}
\usepackage{url}
\usepackage{float}
\usepackage{multirow}
\usepackage{hhline}

\hypersetup{
    pdfnewwindow=true,      
    colorlinks=true,       
    linkcolor=blue,          
    citecolor=blue,        
    filecolor=blue,      
    urlcolor=blue        
}

\newcommand{\orcid}[1]{\begingroup
  \hypersetup{hidelinks}\href{https://orcid.org/#1}{\includegraphics[width=10pt]{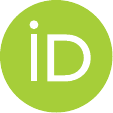}} \endgroup}
  
\begin{document}

\preprint{FERMILAB-PUB-23-515-T}

\title{First Detailed Calculation of Atmospheric Neutrino Foregrounds \\ to the Diffuse Supernova Neutrino Background in Super-Kamiokande}

\author{Bei Zhou \orcid{0000-0003-1600-8835}\,}
\email{beizhou@fnal.gov}
\affiliation{Theoretical Physics Department, Fermi National Accelerator Laboratory, Batavia, Illinois 60510, USA}
\affiliation{William H. Miller III Department of Physics and Astronomy, Johns Hopkins University, Baltimore, Maryland 21218, USA}

\author{John F. Beacom \orcid{0000-0002-0005-2631}\,}
\email{beacom.7@osu.edu}
\affiliation{Center for Cosmology and AstroParticle Physics (CCAPP), Ohio State University, Columbus, Ohio 43210, USA}
\affiliation{Department of Physics, Ohio State University, Columbus, Ohio 43210, USA}
\affiliation{Department of Astronomy, Ohio State University, Columbus, Ohio 43210, USA}

\date{May 6th, 2024}


\begin{abstract}
The Diffuse Supernova Neutrino Background (DSNB) --- a probe of the core-collapse mechanism and the cosmic star-formation history --- has not been detected, but its discovery may be imminent.  A significant obstacle for DSNB detection in Super-Kamiokande (Super-K) is detector backgrounds, especially due to atmospheric neutrinos (more precisely, these are foregrounds), which are not sufficiently understood.  We perform the first detailed theoretical calculations of these foregrounds in the range 16--90 MeV in detected electron energy, taking into account several physical and detector effects, quantifying uncertainties, and comparing our predictions to the 15.9 livetime years of pre-gadolinium data from Super-K stages I--IV.  We show that our modeling reasonably reproduces this low-energy data as well as the usual high-energy atmospheric-neutrino data.  To accelerate progress on detecting the DSNB, we outline key actions to be taken in future theoretical and experimental work.  In a forthcoming paper, we use our modeling to detail how low-energy atmospheric-neutrino events register in Super-K and suggest new cuts to reduce their impact.  
\end{abstract}

\maketitle


\section{Introduction}
\label{sec_introduction}

The Diffuse Supernova Neutrino Background (DSNB) is the flux of all flavors of neutrinos and antineutrinos from massive-star core collapses in cosmic history~\cite{Beacom:2010kk, Horiuchi:2008jz, Jeong:2018yts, DeGouvea:2020ang, Das:2021lcr, Suliga:2021hek, deGouvea:2022dtw, Bell:2022ycf, Akita:2022etk, Suliga:2022ica}.  The first detection and eventual precision measurement of the DSNB will each be of great importance.  While detecting a Milky Way supernova will precisely measure one burst~\cite{Kotake:2005zn, Dasgupta:2009mg, Huedepohl:2009wh, Scholberg:2012id, Adams:2013ana, Jana:2022tsa, Segerlund:2021dfz, Hansen:2019giq, Franarin:2017jnd, Horiuchi:2017qlw, Lu:2016ipr, Fischer:2015oma, Chang:2022aas}, detecting the DSNB will probe the average neutrino emission per core collapse, including from failed, optically dark collapses~\cite{Kochanek:2008mp, Lunardini:2009ya, Hidaka:2015cka, Zaizen:2018wfg, Neustadt:2021jjt, Ivanov:2021lun, Ashida:2022nnv}. And, while the wait for a Milky Way supernova may be long, the DSNB is always present.  The strongest DSNB flux limits~\cite{Malek:2002ns, Super-Kamiokande:2011lwo, Super-Kamiokande:2013ufi, Super-Kamiokande:2021jaq} are from Super-Kamiokande (Super-K), a water-Cherenkov detector with a fiducial volume of 22.5 kton~\cite{Fukuda:2002uc}, which probes DSNB $\bar{\nu}_e$ via inverse beta decay ($\bar{\nu}_e + p \rightarrow e^+ + n$) on free protons~\cite{Vogel:1999zy, Strumia:2003zx}, where the electron (hereafter, we use this to mean an electron or positron, unless we specify otherwise) is detected by its Cherenkov light and the neutron is typically not detected (we focus on the 15.9 livetime years of data from Super-K stages I--IV, before gadolinium was added).

\begin{figure}[b]
\includegraphics[width=0.95\columnwidth]{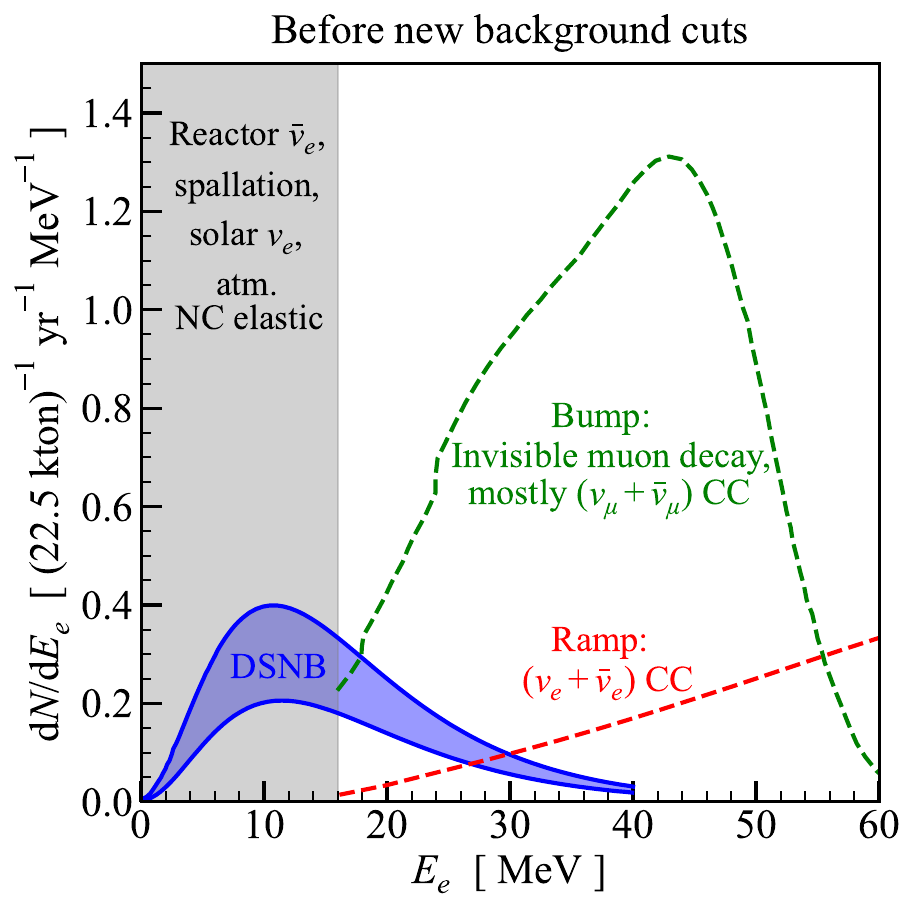}
\caption{Schematic Super-K DSNB search.  The example DSNB signal has a 6-MeV effective temperature after neutrino mixing~\cite{Horiuchi:2008jz}.  The example atmospheric-neutrino backgrounds, which depend on detector energy resolution and detection efficiency, are shown by Super-K's stage-I fits~\cite{Super-Kamiokande:2011lwo, Super-Kamiokande:2021jaq}.}
\label{fig_intro}
\end{figure}

Figure~\ref{fig_intro} illustrates how the DSNB signal is presently obscured by detector backgrounds (e.g., radioactivities induced by muon spallation) and foregrounds (interactions induced by other neutrinos, which we hereafter call backgrounds, following common usage).  The basic physics of these backgrounds is known, but the details are not.  In the present DSNB analysis window, 16--90~MeV in electron total energy $E_e$, the dominant backgrounds are due to atmospheric-neutrino interactions with oxygen nuclei~\cite{Malek:2002ns, Super-Kamiokande:2011lwo, Super-Kamiokande:2013ufi, Super-Kamiokande:2021jaq}; below 16 MeV, several backgrounds contribute~\cite{Li:2014sea, Li:2015kpa, Li:2015lxa, Super-Kamiokande:2015xra, Super-Kamiokande:2019hga, Super-Kamiokande:2021snn, Sakai:2023xdi, Nairat}.  The larger atmospheric component (``the bump") is due to the at-rest decays of invisible (sub-Cherenkov) muons, mostly produced by the charged-current (CC) interactions of ($\nu_\mu + \bar{\nu}_\mu$).  The smaller component (``the ramp") is mostly due to the CC interactions of $\nu_e+ \bar{\nu}_e$.  The DSNB signal and atmospheric-neutrino background rates are $\sim$ 2--4 and $\sim 50$ events/yr in 16--90~MeV.  Better understanding the backgrounds will lead to new cuts to reduce them.

In this paper, we perform the first detailed calculations of atmospheric-neutrino backgrounds for DSNB searches, going well beyond earlier theoretical work~\cite{Beacom:2003nk, Bell:2020rkw} as well as the empirical treatment employed by Super-K in Refs.~\cite{Malek:2002ns, Super-Kamiokande:2011lwo, Super-Kamiokande:2013ufi}, where they fit the normalizations of the backgrounds to data.  More recently, in Ref.~\cite{Super-Kamiokande:2021jaq}, they mention modeling the backgrounds, but almost no details or results are reported.  We model the atmospheric-neutrino fluxes and their mixing, the neutrino-nucleus interactions in water, how the produced particles propagate and register in the detector, and the effects of Super-K data cuts.  We take into account several effects, quantify uncertainties, and compare to data.  We focus on atmospheric-neutrino CC interactions and certain high-energy neutral-current (NC) interactions that produce events in 16--90 MeV, showing that we can also match the data at higher energies.  Our results will also be useful for SK-Gd (with added gadolinium to enhance neutron detection, which began in 2020~\cite{Beacom:2003nk, Super-Kamiokande:2021the, Super-Kamiokande:2023xup}) and Hyper-Kamiokande~\cite{Abe:2018uyc}, as well as other water Cherenkov detectors, e.g., ANNIE~\cite{ANNIE:2015inw, Fischer:2020qht} and WATCHMAN~\cite{Askins:2015bmb}.  

In Sec.~\ref{sec_morebkgd}, we review the broad range of inputs needed for this work.  In Sec.~\ref{sec_repHE}, we show that our calculations reasonably reproduce the most relevant high-energy Super-K atmospheric-neutrino data, an important check.  In Sec.~\ref{sec_repLEinvmu} and Sec.~\ref{sec_repLEnuecc}, we show that our calculations also reasonably reproduce Super-K data in the range 16--90 MeV.  This, along with our detailed accounting of the inputs and uncertainties, is our main result.  In Sec.~\ref{sec_LEparent}, we calculate the parent-particle spectra of the atmospheric-neutrino backgrounds, which provide important physical insights.  In Sec.~\ref{sec_discussion}, we point out key actions to be taken for progress.  We conclude in Sec.~\ref{sec_conclusion}.  In a forthcoming paper~\cite{DSNB2}, we detail how low-energy atmospheric events register in Super-K, which will improve how well they can be identified and controlled.


\section{Framing the Problem}
\label{sec_morebkgd}

In this section, we first review the expected DSNB signal and the observed backgrounds in Super-K.  We then discuss how the atmospheric-neutrino backgrounds are formed, taking into account their fluxes and mixing, their interactions, and how these events register in the detector.  Though many of these points are known, this detailed synthesis is new.


\subsection{Predicted DSNB signals}
\label{sec_morebkgd_sgnl}

The DSNB is a guaranteed flux, time-independent and isotropic, of all neutrino species~\cite{Beacom:2010kk}.  Once detected, it will provide new information on the core-collapse mechanism as well as the cosmic core-collapse rate.  Moreover, as core collapses that directly produce black holes have neutrino signals comparable to or larger than those that produce successful supernovae~\cite{Nakazato:2008vj, Lunardini:2009ya}, the DSNB probes also the rate of failed supernovae~\cite{Kochanek:2008mp, Lien:2010yb, Ashida:2022nnv}.  {\it If the DSNB flux is not found after modest improvements in sensitivity, then there must be surprising new physics or astrophysics.}

As an example DSNB model, we use one of the predictions from Ref.~\cite{Horiuchi:2008jz} as an illustration.  This model has a relatively optimistic choice of spectrum (6-MeV temperature) and an astrophysical normalization that is in good accord with subsequent work~\cite{Lunardini:2012ne, Kresse:2020nto, Horiuchi:2020jnc, Li:2022myd, Libanov:2022yta}. There are three major inputs for the DSNB signal~\cite{Beacom:2010kk}: the neutrino emission per core collapse~\cite{Ott:2017kxl, Roberts:2016lzn, Nakamura:2016kkl}, the cosmic core-collapse rate~\cite{Hopkins:2006bw, Madau:2014bja}, and the physics of detection~\cite{Malek:2002ns, Super-Kamiokande:2011lwo, Super-Kamiokande:2013ufi, Super-Kamiokande:2021jaq}.  The first is the primary observable, as it can only be measured by neutrino experiments; it is the most uncertain.  The second can be determined by electromagnetic observations, and is relatively well known.  The third is very well known.  

Neutrino mixing in the DSNB signal is included implicitly, as we consider only the effective neutrino spectrum outside the supernova, after all mixing effects have occurred (for active neutrinos, no mixing occurs outside the supernova because neutrinos emerge from the dense matter as incoherent mass eigenstates)~\cite{Dighe:1999bi}.  The measured spectrum can thus be directly compared to the SN 1987A data.  It is a separate problem to relate the observed spectra to the initial neutrino spectra inside the proto-neutron star.

The DSNB flux is obtained from a cosmological line-of-sight integral, with the uncertainties due to cosmological parameters being negligible.  The energy-integrated flux is $\sim 10$ cm$^{-2}$ s$^{-1}$ per flavor for nominal models.  The dominant contribution arises around redshift $z \sim 1$, due to the corresponding star-formation rate being $\sim 10$ times larger than the present rate and due to the contribution from higher redshifts being suppressed by the detector analysis threshold being comparable to the neutrino average energy at emission (where $\langle E \rangle \simeq 3 T$, with $T$ the temperature).  Currently, the best prospects for detecting the DSNB are for $\bar{\nu}_e$ in Super-K, due to the large cross section and the huge, low-background detector.  The cross section~\cite{Vogel:1999zy, Strumia:2003zx} is $\sigma(E_\nu) \simeq 10^{-43} {\rm\ cm}^2 (E_{\bar{\nu}_e} - 1.3 {\rm\ MeV} )^2$, with the outgoing electron carrying $E_e \simeq E_{\bar{\nu}_e} - 1.3 {\rm\ MeV}$ and being emitted near-isotropically.


\subsection{Observed backgrounds in Super-K}
\label{sec_morebkgd_snglBkgd}

For Super-K data, we focus on stages I--IV, for which the key DSNB search paper is Ref.~\cite{Super-Kamiokande:2021jaq}, building on prior work in Refs.~\cite{Malek:2002ns, Super-Kamiokande:2011lwo, Super-Kamiokande:2013ufi}.  The energy resolution and detection efficiency vary from stage to stage, which we take into account.  While added gadolinium (relevant to later data) will help with signal selection and background rejection (see Sec.~\ref{sec_morebkgd_SKdetction}), the pure-water data are important because they correspond to 15.9 livetime years.  The Super-K stage IV data (more than 8 livetime years) are especially important because advances in the electronics have allowed Super-K to save detailed information on all photomultiplier (PMT) hits.  These data can thus be reanalyzed as the physical understanding of the signals and backgrounds improves.

Super-K must contend with multiple backgrounds. Below 16 MeV, the backgrounds are due to reactor- and solar-neutrino interactions, spallation beta decays, and atmospheric-neutrino NC interactions.  While their rates are presently large, these backgrounds can be greatly reduced with added gadolinium~\cite{Beacom:2003nk, Super-Kamiokande:2021the, Super-Kamiokande:2023xup} (see Sec.~\ref{sec_morebkgd_SKdetction}).  In 16--90 MeV, while the backgrounds are substantial, they are not prohibitive. It is a remarkable achievement that the rate for these backgrounds is only about one event per week ($\sim 1.5 \times 10^{-6}$~Hz).  For comparison, the low-energy trigger rate is $\sim 10$~Hz, the cosmic-ray muon rate is $\sim 2$~Hz, and the rates of identified solar and higher-energy atmospheric-neutrino events are each $\sim 25$ per day ($\sim 3 \times 10^{-4}$~Hz)~\cite{Fukuda:2002uc, Super-Kamiokande:2005mbp, Super-Kamiokande:2005wtt, Super-Kamiokande:2008ecj, Super-Kamiokande:2010tar, Super-Kamiokande:2015qek, Super-Kamiokande:2016yck}.

Figure~\ref{fig_intro} shows the two components of the observed atmospheric-neutrino background: the bump and the ramp.  The bump component is due to electrons from muon decay at rest, where these muons are produced with kinetic energies below the Cherenkov threshold, and hence are invisible in Super-K (visible muons are easily cut).  These invisible muons are mostly produced by sub-GeV atmospheric ($\nu_\mu + \bar{\nu}_\mu)$ interactions with oxygen and sometimes hydrogen nuclei.  Below, we show that there is also a moderate contribution due to NC interactions of all-flavor neutrinos up to a few GeV.  Once the muons decay, the isotropically emitted electrons constitute a background for the DSNB. While muon decays occur after a delay of a few microseconds, the time of the initial neutrino interaction is unknown unless a nuclear de-excitation gamma ray is produced, in which case the entire event can be recognized as a background and cut.  The muon-decay spectrum shape is well known, but its normalization is not known a priori.  We predict it (see Sec.~\ref{sec_repLEinvmu}).

The ramp component is due to promptly produced electrons from atmospheric $(\nu_e + \bar{\nu}_e)$ CC interactions.  This component is less important than the bump component, and it can be fit from the 60--90 MeV data, which is important for separating the ramp and bump components.  Neither its spectrum shape nor normalization is known a priori, though the shape seems to be simple.  We predict it and the spectrum shape (see Sec.~\ref{sec_repLEnuecc}).


\subsection{Atmospheric-neutrino fluxes and mixing}
\label{sec_morebkgd_atmNu}

To study the backgrounds induced by atmospheric neutrinos, we need the incoming fluxes and their uncertainties.  For the DSNB backgrounds, the most important contributions are from neutrino energies below several GeV, and typically well below this.  Atmospheric neutrinos are created through cosmic-ray interactions in the air, producing secondaries, including pions and kaons, which decay in flight, producing neutrinos~\cite{Gaisser:2002jj, Gaisser:2005dt, Wendell:2015ona}.  In this energy range, the main production mechanism is the decays of pions ($\pi^+ \rightarrow \mu^+ + \nu_\mu$ and its charge conjugate) and muons (from pion decay; $\mu^+ \rightarrow e^+ + \nu_e + \bar{\nu}_\mu$ and its charge conjugate).  At low energies, the flux ratio (combining neutrinos and antineutrinos) is $\nu_e : \nu_\mu : \nu_\tau \simeq 1:2:0 $ before neutrino mixing.  Above several GeV, the $\nu_e$ fraction decreases due to some muons reaching the ground before decaying. 

For the atmospheric-neutrino fluxes, we use those of Ref.~\cite{Honda:2015fha} for $E_\nu > 100$~MeV and those of Ref.~\cite{Battistoni:2005pd} for 10--100~MeV, both evaluated at the location of Super-K.  These join reasonably well; we do not adjust the normalizations so that the size of the mismatch can be judged in the context of the whole spectrum. Given their extensive use, employing these calculations facilitates direct comparison with existing literature. These fluxes are in reasonable accord with those in more recent work, e.g., Ref.~\cite{Zhuang:2021rsg}. Further, as discussed below, while the flux uncertainties are important, they are not limiting. Because we consider a long time period for Super-K stages I--IV, we use the atmospheric-neutrino flux averaged over the solar cycle; in future work, it would be interesting to take the expected flux variations into account.  The neutrino flux uncertainties mostly arise from those of the primary cosmic-ray flux and hadronic secondary production.  To be conservative, the neutrino flux uncertainties that we adopt, following comparisons of the values quoted in Refs.~\cite{Battistoni:2005pd, Honda:2006qj, Barr:2006it}, are $\sim 25\%$ for 10--100~MeV, $\sim 20\%$ for 0.1--1 GeV, and $\sim 15\%$ for 1--10~GeV.  These uncertainties should improve in the near future.  A recent paper, Ref.~\cite{Evans:2016obt}, finds smaller uncertainties in light of new cosmic-ray measurements, but its flux predictions do not yet cover the full energy range we need.

Before entering Super-K, these neutrinos undergo mixing, which depends on energy and baseline (and thus the zenith angle of the neutrino direction).  The dominant effects are due to maximal $\nu_\mu \rightarrow \nu_\tau$ vacuum oscillations with the (large) atmospheric mass-squared splitting, which push the flavor ratios toward $\nu_e : \nu_\mu : \nu_\tau \simeq 1:1:1$, after which oscillations with the (small) solar mass-squared splitting have little effect.   With increasing energy and baseline, Earth matter effects become important.  We calculate the full three-flavor mixing effects (vacuum and matter) using {\tt nuCraft}~\cite{Wallraff:2014qka}, which computes zenith-angle, energy, and production-height dependent mixing probabilities by direct numerical integration.  The uncertainties due to oscillation parameters~\cite{Capozzi:2017ipn, Gariazzo:2018pei} are smaller than those on the flux.


\subsection{Neutrino interactions in water}
\label{sec_morebkgd_nuInt}

For the DSNB signal, which ranges up to a few tens of MeV, the most important interaction is $\bar{\nu}_e + p \rightarrow e^+ + n$ on free (hydrogen) protons.  Neutrino interactions with nuclei are suppressed by binding effects due to the low neutrino energies and interactions with electrons are suppressed by their small masses.  The signal detection cross section is well understood~\cite{Vogel:1999zy, Strumia:2003zx} (see Ref.~\cite{Ricciardi:2022pru} for a more recent calculation).

For atmospheric neutrinos, which range from several tens of MeV to a peak in the GeV range to a falling spectrum at higher energies, the dominant interactions are with bound nucleons in oxygen.  Except at the lowest neutrino energies, interactions with hydrogen are suppressed by the low number density.

In the simplest description, the two most important CC interactions that form backgrounds to the DSNB signal are $\nu_\mu +\,^{16}{\rm O} \rightarrow \mu^- +\,^{15}{\rm O} + p$ and $\bar{\nu}_\mu +\,^{16}{\rm O} \rightarrow \mu^+ +\,^{15}{\rm N} + n$, where these are interactions with bound nucleons that are ejected as they are transformed by the CC weak interaction.  The underlying physics of the initial neutrino-nucleon interaction may be through quasi-elastic scattering (QES), as just described, which is dominant for $E_\nu \lesssim 1$~GeV.  At higher energies, up to $E_\nu \sim$ a few GeV, resonance production (RES), in which the nucleon is briefly excited to a delta resonance, is dominant.  At still higher energies, deep inelastic scattering (DIS), in which the neutrino interacts with quarks, becomes dominant.  These energy ranges indicate approximate guideposts, not sharp separations.  We show below that for the atmospheric-neutrino events we focus on, CCQES interactions are dominant, with some contributions due to NCRES.

In slightly more detail, neutrino-nucleus interactions (both CC and NC) in this energy range are assumed to occur in two steps.  First, a neutrino interacts with a single bound nucleon that has an initial momentum due to its Fermi motion.  Interactions with pairs of nucleons (also called 2p/2h interactions) are also important but are less common.  The interaction with the initial nucleon(s) can be affected by nuclear shadowing, and the availability of final states for the struck nucleon(s) can be affected by Pauli blocking.  Second, the struck (and possibly transformed) nucleon(s), along with any other produced hadrons, travel through the nucleus, where they may interact.  This is called intranuclear hadron transport or final-state interactions and it complicates the whole picture of neutrino-nucleus interactions~\cite{Golan:2012wx, Dytman:2011zz}.

Because neutrino-nucleus interactions are complex and hard to handle analytically, we simulate them using {\tt GENIE}, which provides a comprehensive framework~\cite{Andreopoulos:2009rq, GENIE3_manual, GENIEweb}.  {\tt GENIE} takes into account the neutrino-nucleon/nucleus interaction vertices, nuclear effects, hadronization, final-state interactions, de-excitations of the final-state nucleus, and more, though all approximately. For the 2p/2h interactions, it uses a meson-exchange-current (MEC) model.  A {\tt GENIE} model set is a comprehensive model configuration (CMC) with a specific tune~\cite{GENIE_CMC_tune}. A CMC sets the models and parameters for the above physical aspects, while each CMC has several different tunes, selected to fit varying choices of datasets (neutrino-, electron-, and hadron-nucleus scattering data).  The results from {\tt GENIE} are surely not perfect, but they are adequate to guide our exploration of the physics of the interaction, how to improve cuts, and how to identify where new inputs are needed.  In future work, it would be interesting to compare results using other neutrino-nucleus cross section codes, such as {\tt ACHILLES}~\cite{Isaacson:2022cwh}, {\tt GIBUU}~\cite{Buss:2011mx}, {\tt NEUT}~\cite{NEUT}, {\tt NuWro}~\cite{Golan:2012rfa}, etc.

We use {\tt GENIE (v3.02.02)}~\cite{GENIEweb} with two different model sets.  The first, which is a priori expected to be more accurate, is ``G18\_10a\_02\_11b,'' where ``G18\_10a'' is the CMC name and the remainder is the tune name~\cite{GENIE_3.02.02_xml}.  The ``G18\_10a'' CMC embeds the best theoretical modeling elements implemented in {\tt GENIE} so far~\cite{GENIE3_manual}. It uses the local Fermi gas nuclear model (LFG) and an implementation of the theory calculation in Ref.~\cite{Nieves:2004wx} by Nieves, Amaro, and Valverde (NAV) for the CCQES and CC multi-nucleon processes (with Coulomb corrections included). The empirical {\tt GENIE} MEC model is used for the NC multi-nucleon processes since they are not included in the calculation of Ref.~\cite{Nieves:2004wx}.  Hereafter, we refer to this model set simply as \textbf{LFG-NAV}.

For comparison, we also use the ``G18\_02a\_00\_000'' model set, for which the CMC is based on the default CMC used in {\tt GENIE v2}. It uses a relativistic Fermi gas nuclear model (RFG) modified by Bodek and Ritchie to incorporate short-range nucleon-nucleon correlations~\cite{Bodek:1981wr}. For CCQES, it uses the Llewellyn-Smith model~\cite{LlewellynSmith:1971uhs} (LS; with Coulomb corrections not included). The empirical {\tt GENIE} MEC model is used for both CC and NC multi-nucleon processes; it is known that this makes the cross sections somewhat too large~\cite{Dytman}.  Hereafter, we refer to this model set simply as \textbf{RFG-LS}.

Both CMCs use the Berger-Sehgal model~\cite{Berger:2007rq} for NC and CC resonance production, the Bodek-Yang model~\cite{Bodek:2002vp} for the NC and CC shallow and deep inelastic scattering, the Berger-Sehgal model~\cite{Berger:2008xs} for NC and CC coherent production of pions, the AGKY model~\cite{Yang:2009zx} for hadronization, and the INTRANUKE/hA 2018 model for final-state interactions.  Other important nuclear effects are also included, including Pauli blocking, shadowing, anti-shadowing, EMC effect, de-excitation, etc. For more details, see Refs.~\cite{GENIE3_manual, GENIE_CMC_tune}.

The neutrino-nucleus cross sections relevant to this work have large uncertainties that are not fully quantified, due to deficiencies on both the experimental and theoretical sides.  We assume overall uncertainties of $\sim 20\%$ in this energy range, based on Refs.~\cite{Aharmim:2006wq, Richard:2015aua}.


\subsection{Physics of detection in Super-K}
\label{sec_morebkgd_SKdetction}

Super-K is a cylinder of diameter 39.3~m and height 41.4~m, filled with 50 ktons of ultrapure water~\cite{Fukuda:2002uc}.  The optically isolated inner detector has a mass of 32 kton, and its fiducial volume has a mass of 22.5 kton.  Super-K detects Cherenkov light from relativistic charged particles via $\sim 11,000$ 50-cm PMTs that view the inner detector (the outer detector is viewed by a smaller number of smaller PMTs, providing an active veto layer).  The inner-detector PMTs cover $\simeq 40\%$ (in Super-K stage II, it was $\simeq 20\%$, which worsened the energy resolution) of the detector wall, are sensitive to photons of wavelength $\simeq 300$--700~nm, and have a quantum efficiency of $\sim 0.1$.  Below, we give more details about stage-dependent detector properties such as the energy resolution and efficiency.  Very roughly, at the relevant energies, the energy resolution is $\lesssim 12\%$ ($\lesssim 18\%$ for Super-K stage II) and improving with increasing energy, while the efficiency is $\sim 90\%$ (except for Super-K stage IV at some energies, as discussed in Sec.~\ref{sec_repLEinvmu_calc_threshold}).

We manage most of the detection-related calculations with our own codes.  Where the physics is more complex, we simulate or check our results using {\tt FLUKA}~\cite{Ferrari:2005zk}, which provides a comprehensive framework for particle transport in varying media, covering particle creation, interactions and energy loss, decay and capture, and more.  Here, our main use of {\tt FLUKA} is the propagation of nuclear gamma rays, charged pions, and muons from neutrino interactions in water.  In Ref.~\cite{DSNB2}, where we detail detection signatures, we use {\tt FLUKA} extensively.

To emit Cherenkov radiation, a charged particle's speed must exceed the phase velocity of light in water, which is $c/n$, where $n \simeq 1.33$ is the refractive index of water and $c$ is the speed of light in vacuum.  This sets a theoretical threshold for the lowest-speed particles that can be detected, $\beta_{th} = 1/n \simeq 0.75$, or a Lorentz factor $\gamma_{th} \simeq 1.51$.  For electrons, muons, pions, and protons, this corresponds to kinetic energies of 0.26, 54, 72, and 481 MeV, respectively.  In practice, to be detectable, the kinetic energies must be somewhat higher, as discussed in Sec.~\ref{sec_repLEinvmu_calc_threshold}.  The Cherenkov angle, $\theta_c$, is $\cos{\theta_c} = 1/n\beta$.  For ultrarelativistic particles, $\theta_c = 42^\circ$ is constant, while the angle is less for less-relativistic particles.

The pattern of PMT hits from a particle --- ideally, forming a clear ring-like pattern --- gives information on its type, position, energy, and direction.  Its charge cannot be determined from the Cherenkov light alone.  To set some relevant scales, for relativistic electrons, the light yield corresponds to about 6 detected photoelectrons per MeV and the threshold for solar-neutrino searches is $E_e \simeq 4$ MeV.  For relativistic charged particles at higher energies, the efficiency of detection is nearly perfect (this may be lowered by analysis cuts) and particle identification is very good. 

Some nonrelativistic charged particles are detectable. For example, while sub-Cherenkov muons and charged pions are themselves invisible, their decays or nuclear captures almost always produce detectable signals.  Also, the initial production of sub-Cherenkov muons or charged pions can be accompanied by nuclear gamma rays or neutrons.  And even though neutral particles are invisible themselves, they can lead to detectable signals.  Gamma rays produce detectable electrons through Compton scattering and pair production.  Neutral pions quickly decay to gamma rays. Neutrons produce gamma rays through inelastic interactions with nuclei or their eventual captures on nuclei.  In addition, unstable nuclei that later beta decay can be produced at the initial vertex or through particle propagation.  Further details are discussed in our forthcoming paper~\cite{DSNB2}.

The physical basis we use to separate DSNB signals from atmospheric-neutrino backgrounds is that DSNB $\bar{\nu}_e$ events produce only one electron and one neutron (the latter only detected efficiently in the presence of gadolinium), while atmospheric-neutrino events often produce different final states.  The DSNB event sample is defined by several key cuts in Super-K analyses, which we follow as closely as possible when simulating neutrino interactions with {\tt GENIE} and final-state particle transport with {\tt FLUKA}.  Events must be contained in the fiducial volume with no activity outside.  The Cherenkov radiation should be only one ring, which is fuzzy due to being caused by an electron as opposed to other particles, and its angle should be about $42^\circ$.  For the time structure, there should be only one peak (due to the decay electron), with no other activity for tens of microseconds before or after.  These considerations, along with related ones for higher-energy events, help us define event classes to reproduce the Super-K data (Secs.~\ref{sec_repHE}, \ref{sec_repLEinvmu}, and~\ref{sec_repLEnuecc}), and identify new ways to reduce backgrounds~\cite{DSNB2}.

Though we focus on the pure-water data from Super-K in stages I--IV, we note that starting in 2020, Super-K began upgrading to SK-Gd by adding dissolved gadolinium sulfate, Gd$_2$(SO$_4$)$_3$~\cite{Beacom:2003nk, Super-Kamiokande:2021the, Super-Kamiokande:2023xup}.  In 2020, the concentration of gadolinium by mass was set to 0.01\%, and in 2022, it was increased to 0.03\%; it may ultimately reach the originally proposed target of $0.1\%$, near which 90\% of neutrons capture on gadolinium~\cite{Beacom:2003nk}.  In pure-water data, a neutron captures on a free proton, releasing a 2.2-MeV gamma ray~\cite{nndcweb}, which is hard to detect~\cite{Super-Kamiokande:2013ufi, Super-Kamiokande:2021jaq}.  With gadolinium, which has a huge cross section for thermal-neutron capture, the energy release is a total of $\simeq 8$ MeV in a few gamma rays~\cite{nndcweb}, which is easy to detect.  The ability to tag neutrons identifies the DSNB signal (with one neutron) and rejects many backgrounds that have zero or multiple neutrons~\cite{Beacom:2003nk, Super-Kamiokande:2021the, Super-Kamiokande:2023xup}.  Further progress on background rejection will come from improved spallation cuts~\cite{Li:2014sea, Li:2015kpa, Li:2015lxa, Super-Kamiokande:2015xra, Super-Kamiokande:2021snn, Nairat} and better reconstruction of atmospheric-neutrino NC events~\cite{Super-Kamiokande:2019hga, Sakai:2023xdi}.  The long-term goal is to suppress the background down to $\simeq 10$ MeV, below which reactor neutrino backgrounds will remain overwhelming.


\section{Calculation Validation by Matching Super-K High-Energy Atmospheric-Neutrino Data}
\label{sec_repHE}

In this section, we show that our calculations reasonably reproduce the most relevant GeV-range Super-K atmospheric-neutrino data~\cite{Ashie:2005ik}.  This is a precondition to accurately predicting the low-energy (16--90 MeV) backgrounds for DSNB searches.

For the high-energy comparison, we use electron-like and muon-like data from Super-K stage I, spanning from April 1996 to July 2001 (4.1 years livetime, 22.5-kton fiducial volume)~\cite{Ashie:2005ik}.  These data are nearly ideal: the detection efficiency is near unity (in contrast to the low-energy data in the next sections), the momentum and angular resolution values are smaller than Super-K's bin widths, the particle identification is very good, and backgrounds are negligible.  In the energy range we focus on, the physical final states and detectable event classes are relatively simple.  Because our focus is on neutrino-nucleus interactions, we use Super-K's angle-averaged spectra; the zenith-angle distributions mostly test neutrino mixing.  
For the Super-K data, there is a more recent paper with a much larger exposure (stage I to IV)~\cite{Richard:2015aua}, but it does not provide the charged-lepton spectra we need, instead providing only model-dependent reconstructed neutrino spectra (which our results agree with; not shown).

We focus on comparing to the angle-averaged momentum spectra of Super-K's fully contained (FC) single-ring events, which dominate below about 1 GeV~\cite{Ashie:2005ik}.  First, this energy range is closest to what we need for the calculations in Sec.~\ref{sec_repLEinvmu} and Sec.~\ref{sec_repLEnuecc}.  Second, the distances that the charged leptons travel are small compared to the detector’s size (e.g., muons travel only $\simeq 5\, {\rm m}\, (E / 1\, \rm GeV)$ and electrons much less), so that the detector geometry can be ignored.  For ($\nu_e + \bar{\nu}_e$), the Super-K FC event spectra start at a momentum of 0.1 GeV, while for ($\nu_\mu + \bar{\nu}_\mu$), they start at a momentum of 0.18 GeV.

The momentum spectrum of a particle $f$ can be calculated by summing the interaction channels over neutrino species, $\nu$, and targets, $T$,
\begin{multline}
\frac{d N_f}{d p_f} =
\Delta t
\sum_{\nu T \rightarrow f} N_{T} 
\int dE_\nu\, \frac{d\sigma_{\nu T \rightarrow f}}{d p_f}(E_\nu,p_f) \\
\int d\Omega_z \, 
\frac{d\Phi}{dE_\nu}(E_\nu, \cos\theta_z, \phi_z) \, P_{\rm osc}(E_\nu, \cos\theta_z)\, ,
\label{eq_calc_HE}
\end{multline}
where $\Delta t$ is the exposure time, $N_T$ the number of targets in the fiducial volume ($N_{\rm O} = 7.5 \times 10^{32}$, $N_{\rm H} = 1.5 \times 10^{33}$, and $N_e = 7.5 \times 10^{33}$, though interactions with electrons are negligible), and $\Omega_z(\cos{\theta_z}, \phi_z)$ is the solid angle defined with the zenith as the axis direction.  The initial atmospheric-neutrino flux is ${d\Phi}/{dE_\nu}$, where we ignore absorption in Earth due to the low neutrino energies.  The oscillation probability is $P_{\rm osc}(E_\nu, \cos\theta_z)$; the convolution of this with the flux over angles gives the neutrino flux after mixing.  Then convolution with the differential cross section, ${d\sigma_{\nu T \rightarrow f}}/{d p_f}$, gives our prediction for the angle-averaged detected event spectrum.

We calculate the differential CC cross sections using {\tt GENIE} simulations.  For simplicity, here we calculate these without regard to Super-K's event classes, meaning that we capture the full CC neutrino-interaction rate (NC contamination is minimal).  \textit{Therefore, at sufficiently high energies, our prediction (for the total spectrum) will exceed the Super-K data (where we select the FC events only).}  For the event classes we neglect, these start to become important above 1 GeV.  For both electron-like and muon-like events, there are contributions from multi-ring events, for which only the total visible energy is measured.  And for muon-like events only, there is also a contribution from partially contained events, for which only the total contained energy is measured.  As noted below, we show the energy ranges where our results should be accurate.

\begin{figure*}[t]
\includegraphics[width=0.93\columnwidth]{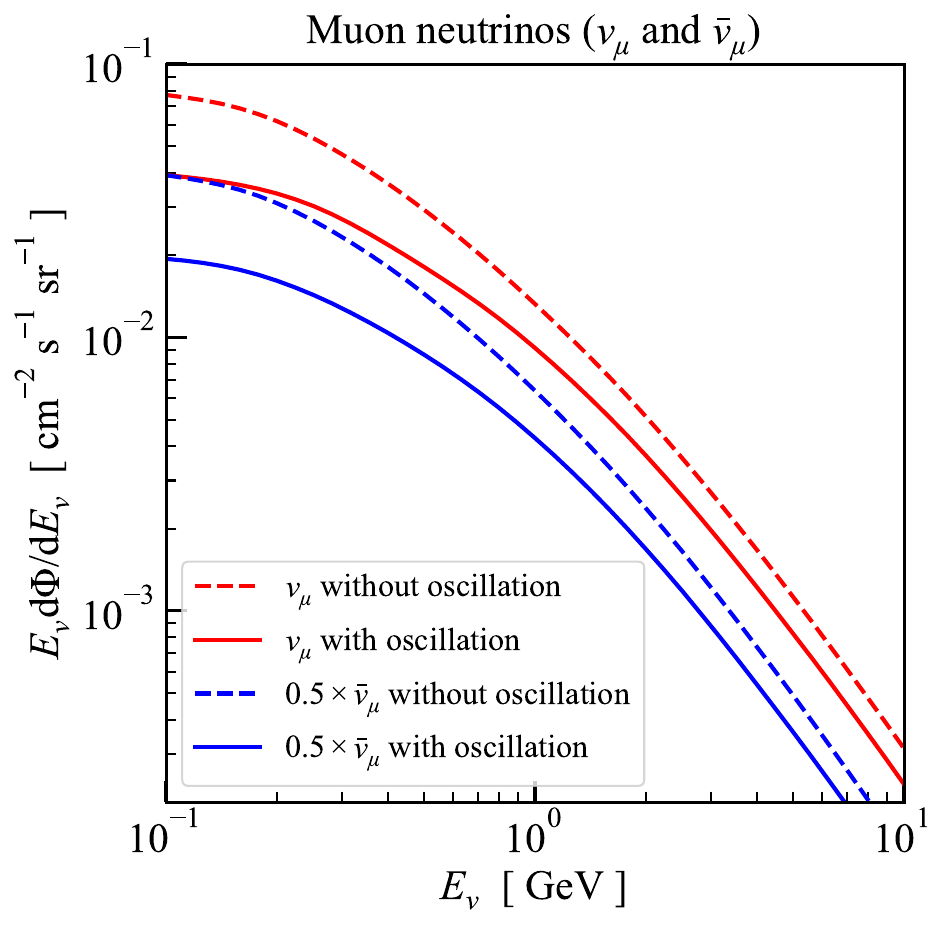}
\includegraphics[width=0.93\columnwidth]{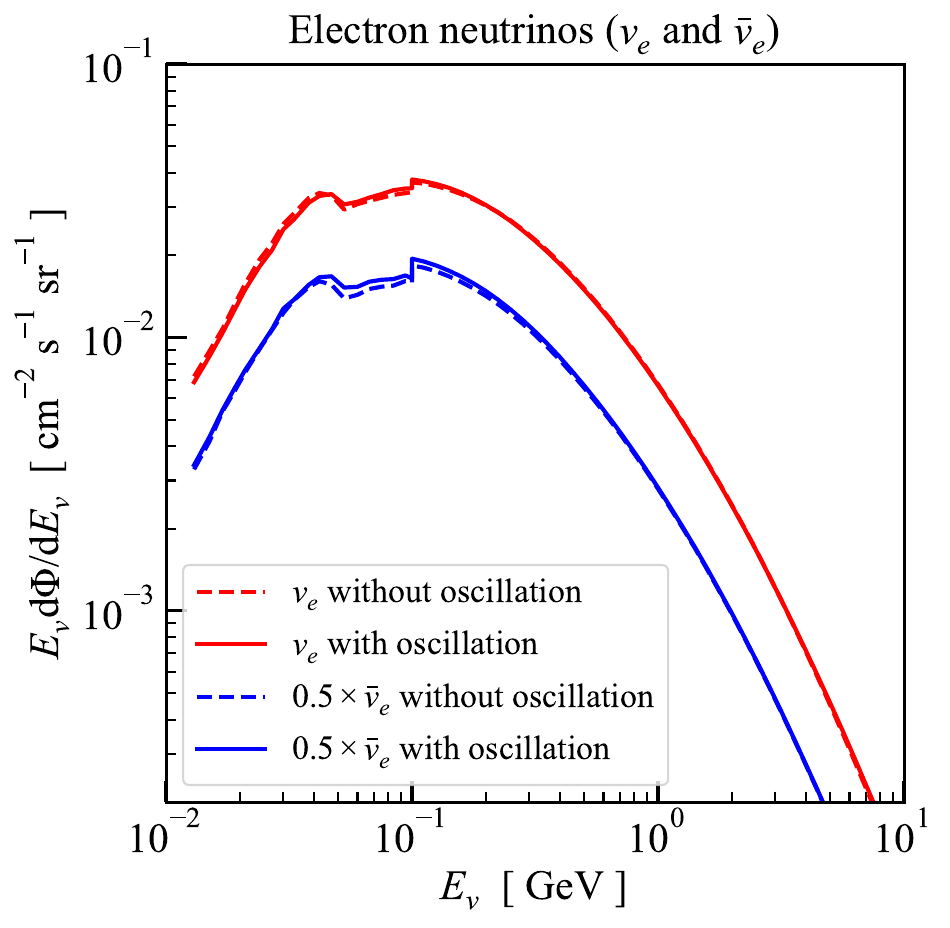}
\caption{Angle-averaged fluxes of atmospheric neutrinos, without and with neutrino mixing, following Refs.~\cite{Battistoni:2005pd, Honda:2015fha, Wallraff:2014qka}.  We have smoothed out some small wiggles due to the discreteness of the HKKM tables.
{\bf Left:} Results for $\nu_\mu$ and $\bar{\nu}_\mu$.  {\bf Right:} Results for $\nu_e$ and $\bar{\nu}_e$ (note the change in the energy range; the notch at 0.1~GeV is due to joining different flux predictions; the bump around 45~MeV is due to the Michel spectrum from muon decay at rest). For $\bar{\nu}$ curves in both panels, we have multiplied the fluxes by 0.5 so that they do not overlap with the $\nu$ curves.
}
\label{fig_atm_flux}
\end{figure*}

\begin{figure*}
\includegraphics[width=0.94\columnwidth]{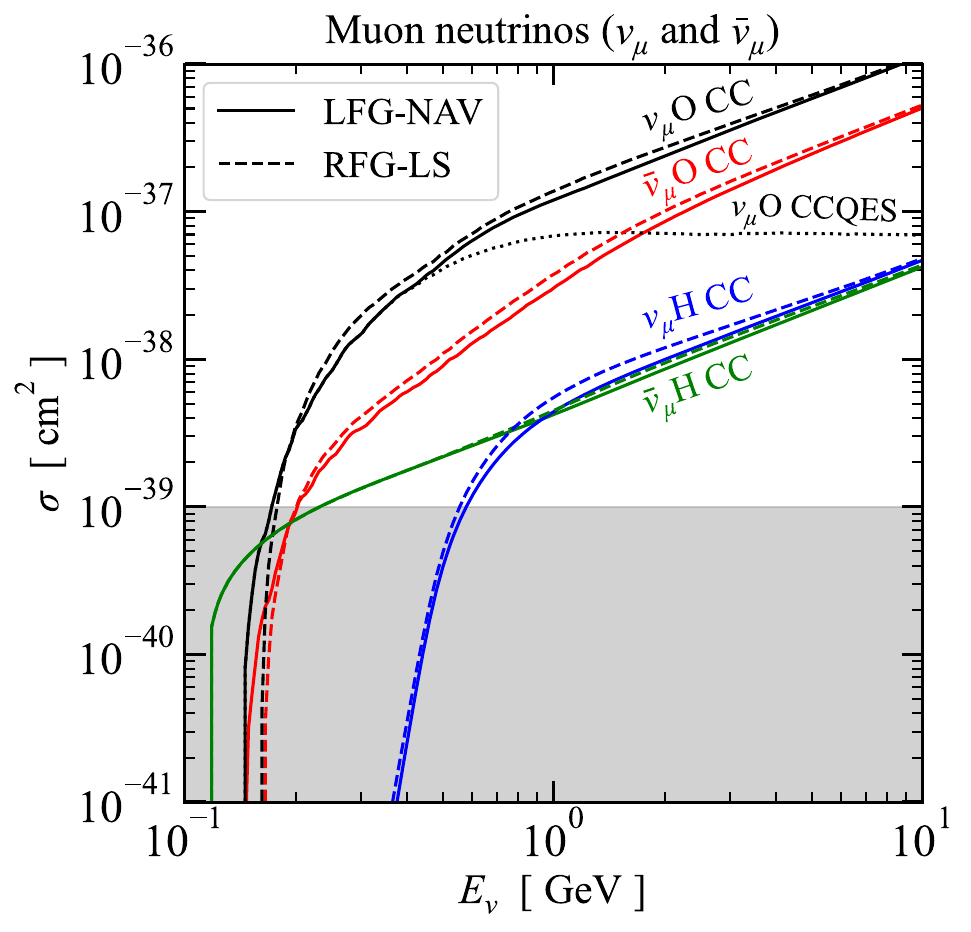}
\includegraphics[width=0.94\columnwidth]{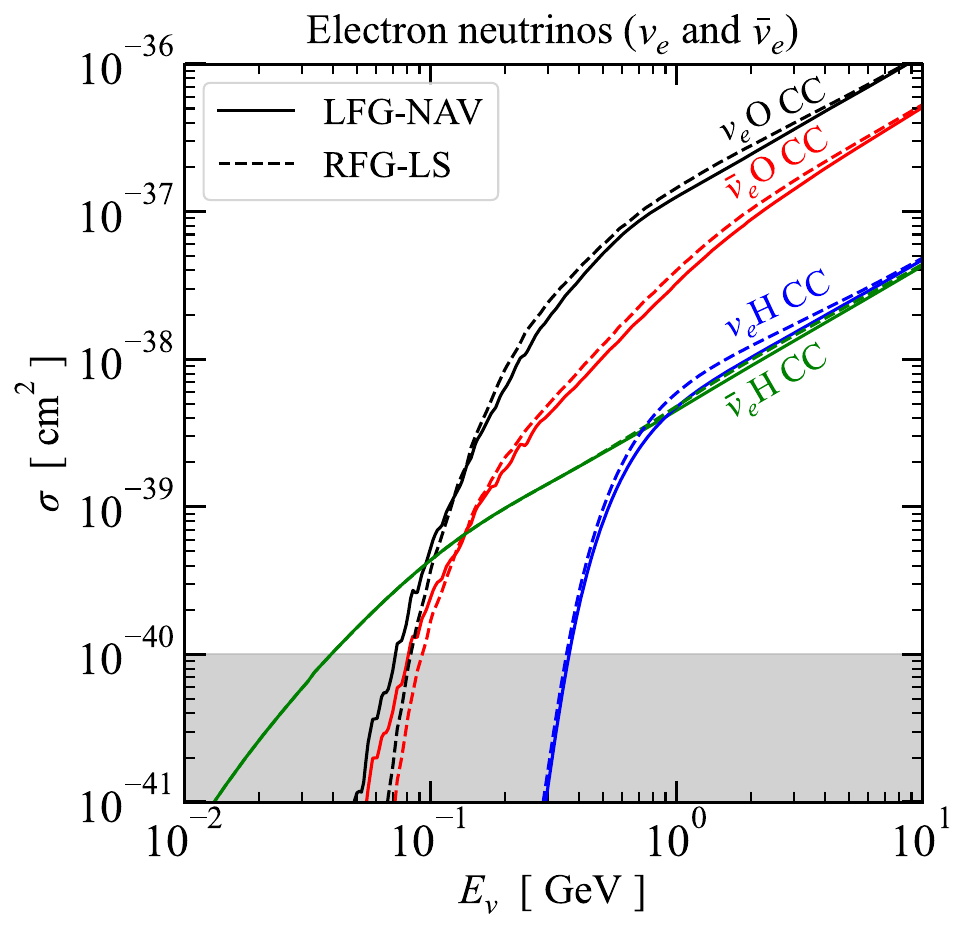}
\caption{{\tt GENIE} cross sections~\cite{GENIE_3.02.02_xml} for the most important CC interactions, showing results for the two model sets used in our work (solid lines for the LFG-NAV model set and dashed lines for the RFG-LS model set).  {\bf Left:} Results for $\nu_\mu$ and $\bar{\nu}_\mu$.  {\bf Right:} Results for $\nu_e$ and $\bar{\nu}_e$ (note the change in the energy range). Below $\sim 1$ GeV, CCQES dominates, as shown for one specific channel ($\nu_\mu$O CCQES) in the left panel.  As expected, the cross sections are somewhat larger for the RFG-LS model set.  The gray shading roughly indicates the neutrino-oxygen cross section ranges that are too small to affect our results (see Figs.~\ref{fig_parent_spec_invmu_LFG} and \ref{fig_parent_spec_nuecc_LFG}).  For the neutrino-hydrogen cross section, which is more certain, somewhat smaller cross section values are relevant.
} 
\label{fig_xsec}
\end{figure*}

\begin{figure*}
  \begin{minipage}{\textwidth}
    \includegraphics[width=0.49\textwidth]{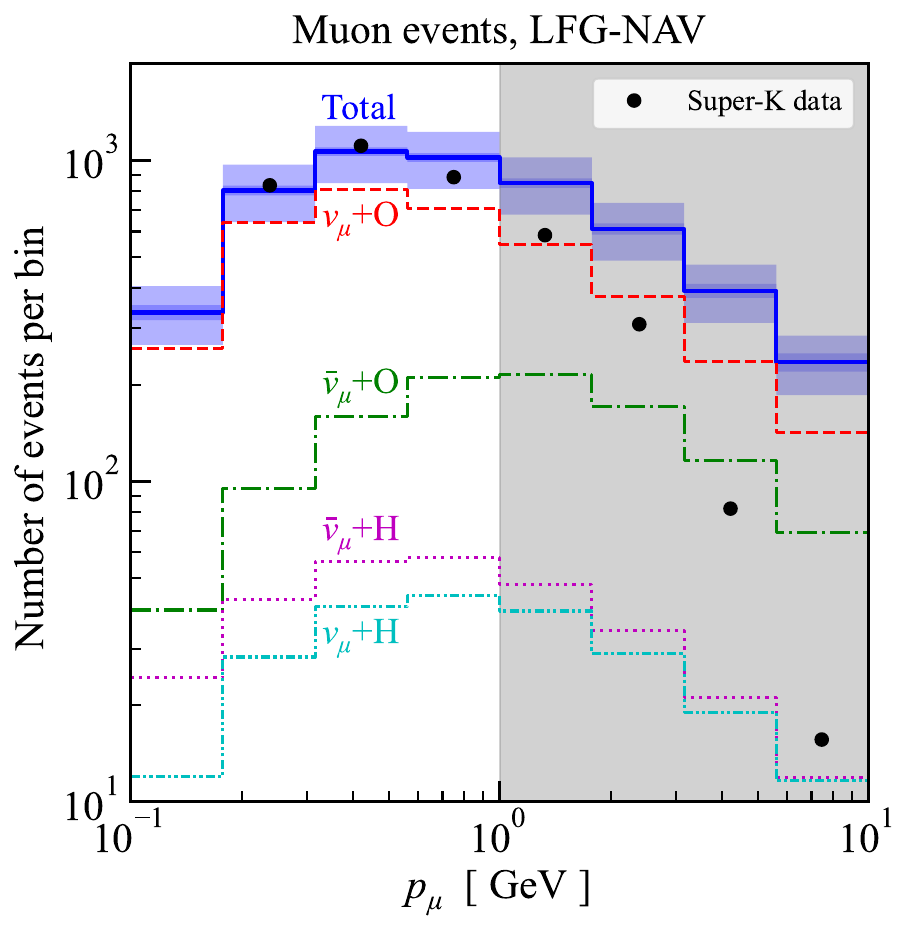}
    \includegraphics[width=0.49\textwidth]{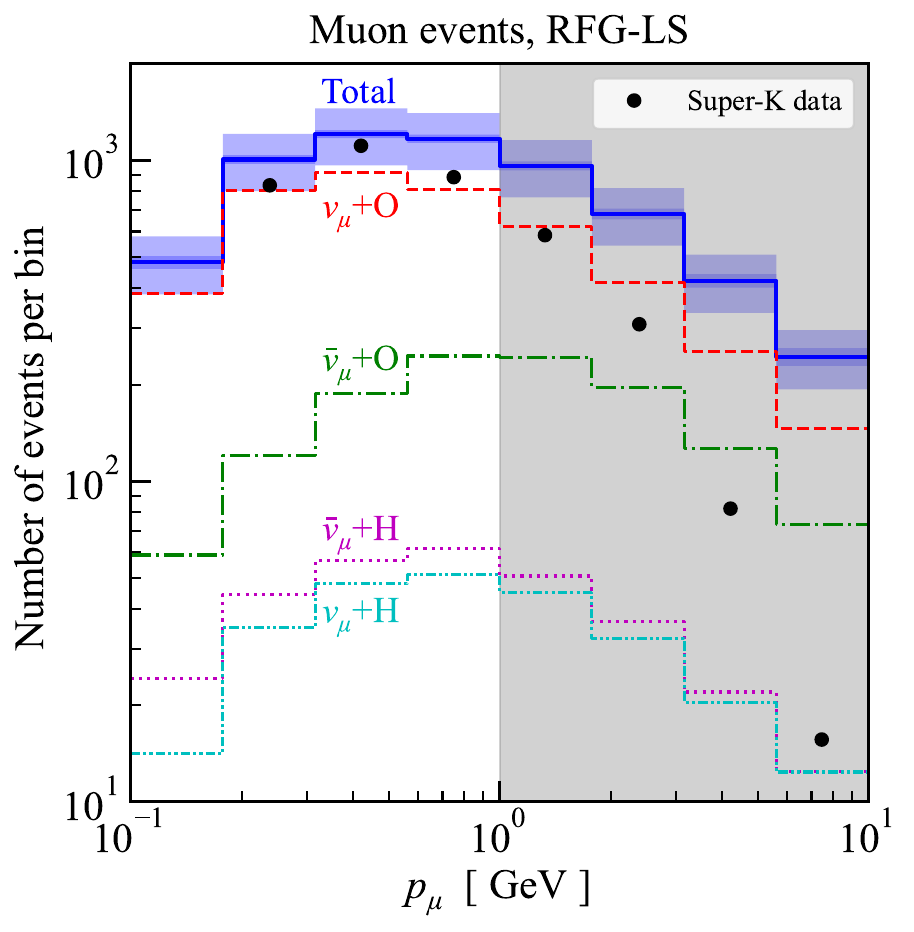}
  \end{minipage}
  \vspace{0.2cm}
  \begin{minipage}{\textwidth}
    \includegraphics[width=0.49\textwidth]{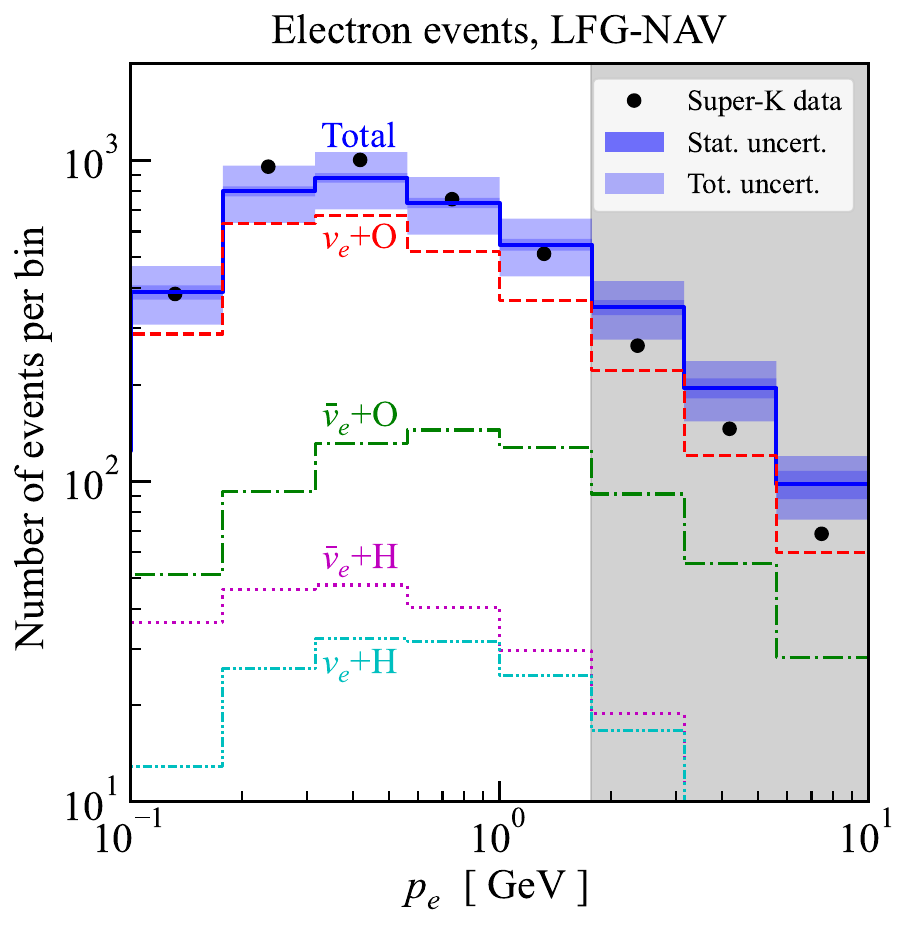}
    \includegraphics[width=0.49\textwidth]{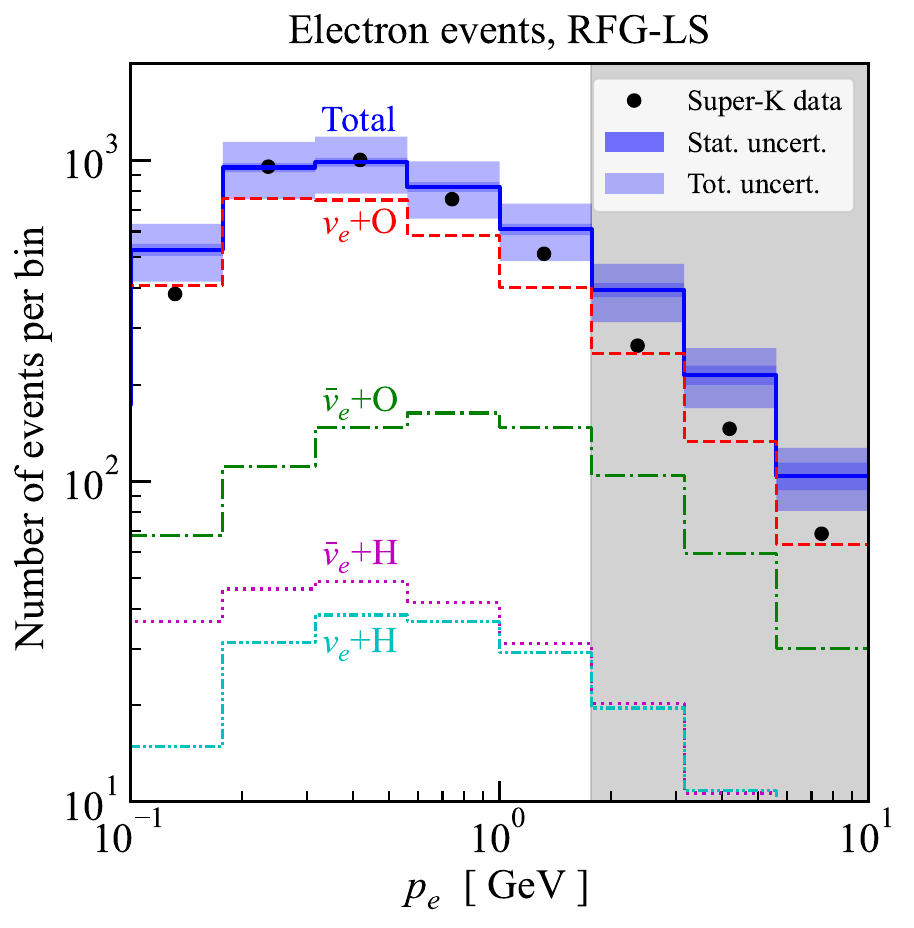}
  \end{minipage}
\caption{Our calculated results for the charged-lepton spectra induced in Super-K stage I by atmospheric neutrinos, compared to measurements (points)~\cite{Ashie:2005ik}, with the statistical and total uncertainties (i.e., statistical and systematic uncertainties added in quadrature) shown in the bands.  {\bf Upper:} Muon-like FC single-ring data compared to our calculations for the total rate of all event classes.  In the gray region, our neglect of other event classes in the data means that our predictions should be increasingly too large.  {\bf Lower:} Same for electron-like events.  \textbf{Left:} Results for the LFG-NAV model set. \textbf{Right:} Results for the RFG-LS model set.  Both predictions agree reasonably well with data, though the RFG-LS predictions are somewhat too large at low energies, as expected.}
\label{fig_reproduce_HE}
\end{figure*}

Figure~\ref{fig_atm_flux} shows our calculated angle-averaged atmospheric-neutrino fluxes without and with mixing (see Sec.~\ref{sec_morebkgd_atmNu}).  The spectra are plotted as $E dN/dE = (2.3)^{-1} dN/d\log_{10}E$, matching the log scales on the x-axes, so that relative heights of the curves at different energies faithfully show their relative contributions to the integrated flux.  The peaks near 0.1 GeV in neutrino energy follow from the peak near 1 GeV in the cosmic-ray spectrum and the kinematics of pion production and decay. At high energies, the spectra of $\nu_\mu$ and $\bar{\nu}_\mu$ follow the well-known power law for the parent cosmic rays, $dN/dE \sim E^{-2.7}$, while the spectra of $\nu_e$ and $\bar{\nu}_e$ are steeper due to some muons reaching the ground before decaying.

Figure~\ref{fig_xsec} shows the most important CC cross sections, which we obtain from pre-computed tables for {\tt GENIE}~\cite{GENIE_3.02.02_xml}.  In the high-energy range, where the cross sections rise linearly, they are determined primarily by particle-physics considerations.  In contrast, at lower energies, the steeper slope indicates the importance of nuclear effects and hence greater uncertainties.  This is especially prominent in the $\nu_e$ and $\bar{\nu}_e$ cases, where effects due to the charged-lepton mass are negligible.  For nuclear targets, antineutrinos have smaller cross sections than neutrinos, in part due to the cancellation between the vector and axial interactions~\cite{LlewellynSmith:1971uhs}; note that hydrogen targets are a special case due to the lack of neutrons. Below about 1~GeV, QES dominates; starting at few GeV, RES dominates because the thresholds for resonances are surpassed; and above several GeV, DIS dominates because the nucleons are resolved.  The most important target is oxygen, due to its large number of nucleons compared to hydrogen; this changes only at the lowest energies, due to kinematic effects caused by the nuclear binding.

Figure~\ref{fig_reproduce_HE} shows our calculated charged-lepton spectra for Super-K.  The spectra are peaked, due to the convolution of the falling spectra (Fig.~\ref{fig_atm_flux}) and the rising cross sections (Fig.~\ref{fig_xsec}).  In the shaded regions, our predictions are increasingly larger than the data because of, as discussed above, increasingly larger contributions from event classes besides our focused FC single-ring events, including multi-ring events and partially contained events. In the non-shaded regions, the agreement for the LFG-NAV model set is within about ten percent, even better than predicted from uncertainties on the fluxes and cross sections. As expected, the RFG-LS model set gives a predicted spectrum that is somewhat too high, especially at the lowest energies, which is relevant to the calculations in Sec.~\ref{sec_repLEinvmu} and Sec.~\ref{sec_repLEnuecc}.

Note that in Fig.~\ref{fig_reproduce_HE}, we do not show uncertainties on the Super-K data points.  Instead, we show the full uncertainties on our predictions.  The reason is that we want to compare specific models to measured data, as opposed to using the measured data to show the allowed range of models. In other figures below, we take a similar approach.  

The results of this section thus show that the framework we use --- neutrino flux, mixing, total and differential cross sections, and corresponding detector response --- is adequate to reproduce the most relevant high-energy Super-K atmospheric-neutrino data (i.e., those in Fig.~\ref{fig_reproduce_HE} from 0.1 to 1~GeV).  This increases our confidence in the results of the next section, where we continue to lower energies.


\clearpage

\section{New Results on Super-K Low-Energy Atmospheric data: Invisible-Muon Component}
\label{sec_repLEinvmu}

In this section, we calculate our predictions for the invisible-muon component (the bump) of the atmospheric-neutrino background, the larger of the two components.  Previously, in Refs.~\cite{Malek:2002ns, Super-Kamiokande:2011lwo, Super-Kamiokande:2013ufi}, the spectrum shape was predicted but the normalization was fit to data.  More recently, in Ref.~\cite{Super-Kamiokande:2021jaq}, it is mentioned that the measured rate in 29.5--49.5 MeV is 88.8\% as large as predicted, but no details or other results are given.  The decay electrons from invisible muon decays follow the Michel spectrum, with a small distortion due to $\mu^-$ always undergoing atomic capture (mostly on oxygen) and decaying in orbit; the complete spectrum is well measured by Super-K using stopped cosmic-ray muons.  (As discussed below, some $\mu^-$ also undergo nuclear capture, which we take into account.)  \textit{When counting Michel electrons, we always mean in the energy range of 16--90 MeV.}  
We assume that the electrons from muon decay at rest are emitted isotropically because the muons' initial directions are nearly isotropic.
Table~\ref{tab_invmu_SK1234} and Table~\ref{tab_invmu_SK4} summarize all of our predictions.  This is the first time that all of the inputs have been systematically addressed.

As noted, the DSNB signal consists of a single low-energy electron and no other measured detector activity before or after (in pure water, the probability of detecting the neutron capture is low~\cite{Super-Kamiokande:2013ufi, Super-Kamiokande:2021jaq}).  This simplicity makes it straightforward for Super-K to cut backgrounds.


\subsection{Overview of the calculations}
\label{sec_repLEinvmu_calc}

We calculate the muon initial momentum spectra using a formula similar to Eq.~(\ref{eq_calc_HE}), but with some differences:
\begin{widetext}
\begin{equation}
\frac{d N_f}{d p_f} =
\Delta t
\sum_{\nu T \rightarrow f} N_{T}
\int dE_\nu\, \frac{d\sigma_{\nu T \rightarrow f}}{d p_f}(E_\nu,p_f) \otimes Cuts \otimes Corr \otimes Det \,
\int d{\Omega_z} \, 
\frac{d\Phi}{dE_\nu}(E_\nu, \cos\theta_z, \phi_z) P_{\rm osc}(E_\nu, \cos\theta_z)\,.
\label{eq_calc_LE}
\end{equation}
\end{widetext}
In this subsection, we explain the differences.  First, there is one important difference in $\frac{d\sigma_{\nu T \rightarrow f}}{d p_f}(E_\nu,p_f)$ for low-energy atmospheric-neutrino events (Sec.~\ref{sec_repLEinvmu_calc_GENIEchoices}).  We then cover the three new terms:  ``$Cuts$'' means the event classes defined from physically interpreting the Super-K analysis cuts (Sec.~\ref{sec_repLEinvmu_calc_eventClasses}), ``$Corr$'' means several required physical corrections (Sec.~\ref{sec_repLEinvmu_calc_PhysCorr}), and ``$Det$'' means detection effects (Sec.~\ref{sec_repLEinvmu_calc_threshold}). 


\subsubsection{Basic {\tt GENIE} results}
\label{sec_repLEinvmu_calc_GENIEchoices}

In neutrino-nucleus interactions, the residual nucleus is often left in an MeV-range excited state whose decays include prompt gamma-ray emission.  In water, these gamma rays deposit their energy via multiple Compton scattering and sometimes pair production.  Through {\tt FLUKA simulations}, we find that the total Cherenkov yield is a distribution, but that the equivalent electron energy is almost always greater than $0.75 E_\gamma$.  Given the typical nuclear gamma-ray energies here (nearly all above 5 MeV), nearly all of them should produce events in the energy range where Super-K's efficiency is high (e.g., as in their solar-neutrino searches~\cite{Super-Kamiokande:2005wtt, Super-Kamiokande:2008ecj, Super-Kamiokande:2010tar, Super-Kamiokande:2016yck}).  In neutrino interactions that produce relativistic charged particles, this additional Cherenkov light is negligible, but it is important if the event has no other prompt signals, e.g., as in invisible-muon production by atmospheric neutrinos.  When a nuclear gamma ray is detected a few microseconds before an electron, the event can be recognized as an invisible-muon background and rejected.  There are uncertainties in both $Br_\gamma$ (the probability of emitting a nuclear gamma ray) and $\epsilon_\gamma$ (the probability of it leading to a detectable signal in Super-K) that are discussed further below.

For neutrino-oxygen interactions, {\tt GENIE} models the nuclear gamma-ray emission based on Refs.~\cite{Ejiri:1993rh, Kobayashi:2005ut}, for which $Br_\gamma \simeq 50\%$.  Ref.~\cite{Ejiri:1993rh} is a theoretical calculation based on the nuclear shell model, and it gives the nuclear gamma-ray energies and probabilities for the one-nucleon-hole $p_{1/2}$, $p_{3/2}$, and $s_{1/2}$ states reached via interactions with single nucleons. Where present, the emitted nuclear gamma rays are mostly above 6 MeV.  Ref.~\cite{Kobayashi:2005ut} is an experimental measurement of the gamma-ray energies and probabilities for the specific case of the one-proton-hole $s_{1/2}$ state, which corresponds to a higher-energy nuclear excitation.  While the probabilities of reaching these excitations are low (a few percent), the measured data are consistent with $Br_\gamma \simeq 50\%$~\cite{Ejiri:1993rh, Kobayashi:2005ut}.  More recently, another theoretical calculation~\cite{Ankowski:2011ei} and an in-situ experimental measurement in T2K~\cite{T2K:2014vog, T2K:2019zqh} also both give results consistent with $Br_\gamma \simeq 50\%$. 

In {\tt GENIE3} compared to {\tt GENIE2}, MEC models to account for 2p/2h interactions were added to both the LFG-NAV and RFG-LS model sets. These 2p/2h interactions should also lead to final states with nuclear gamma rays, but {\tt GENIE} does not include them.  While Refs.~\cite{Ejiri:1993rh, Kobayashi:2005ut} consider only neutrino interactions with single nucleons (and not 2p/2h interactions), they do provide gamma-ray energies and emission probabilities for several cases where multiple nucleons are ejected due to the high nuclear excitations.  The gamma-ray energies and probabilities should be similar regardless of the reason for ejecting multiple nucleons.  Therefore, following those references, we assume that $\simeq 50\%$ of the MEC events have nuclear gamma-ray emission with energies above 5 MeV. This correction is moderately important for the LFG-NAV model set, as $\simeq15\%$ of the invisible-muon events are due to the MEC component.  It is more important for the RFG-LS model set, where the fraction is instead $\simeq30\%$ (as noted above, the MEC component for the RFG-LS model set is known to be too large~\cite{Dytman}).


\subsubsection{Cuts to {\tt GENIE} results}
\label{sec_repLEinvmu_calc_eventClasses}

We apply cuts to our simulation results (``$Cuts$'' in Eq.~(\ref{eq_calc_LE})) to mimic the many cuts that Super-K applies to real data to isolate the DSNB signal.  To define our background sample of invisible-muon events, the most important particles to keep track of are charged leptons, all types of pions, and nuclear gamma rays.  We also cut other particles, like relativistic protons, but those are rare ($< 1\%$ of events).  To begin, we cut all events where the detectable energy or number of detectable particles is incompatible with the simple, low-energy DSNB signal.

For ($\nu_\mu + \bar{\nu}_\mu$) CC events, we then cut those where the muon is above its theoretical Cherenkov threshold ($p_\mu \simeq 120$~MeV).  Muons with Cherenkov radiation will easily trigger the detector, typically giving a muon-like Cherenkov ring and also a decay electron a few microseconds later.  Such events will be removed by Super-K's cuts.  We call our results at this step our ``Naive'' calculation.  Below, we discuss how we must correct for muons that are not actually detectable because they are not far enough above the Cherenkov threshold.

When an invisible muon is produced, it could be accompanied by one or more pions, though the latter is rare.  We cut all such events.  For a $\pi^0$, it decays to two gamma rays, emitted in nearly opposite directions, due to the modest boosts, and these gamma rays pair-produce or Compton-scatter electrons. This causes two or more rings, allowing for these events to be cut.  For a $\pi^+$ or a $\pi^-$ above the Cherenkov threshold, it will have Cherenkov radiation from itself and from its decay muon and then electron, which makes such events easy to identify.  For a $\pi^+$ below the Cherenkov threshold, it will be cut due to the extra electron from its decay chain.  A $\pi^-$ below the Cherenkov threshold mostly undergoes atomic capture and then nuclear capture~\cite{Ponomarev:1973ya, Measday:2001yr, Czarnecki:2014cxa}.  These usually but not always give detectable signatures but, in any case, the total rate of such events is small.

Events with nuclear gamma rays are cut due to their ``2-peak'' timing features~\cite{Super-Kamiokande:2011lwo}, since these gamma rays are prompt after the neutrino interactions, whereas the Michel electrons are delayed by the muon decay time.  We cut all {\tt GENIE} invisible-muon events with nuclear gamma rays, plus appropriate fractions of the MEC events.  We do not know the efficiency, $\epsilon_\gamma$, of Super-K cutting these gamma rays, as there is no clear documentation about this. In our calculation, we show results for both $\epsilon_\gamma = 0\%$ and $\epsilon_\gamma = 100\%$, then discuss what the observed data suggest.


\subsubsection{Physical corrections to {\tt GENIE} results}
\label{sec_repLEinvmu_calc_PhysCorr}

There are three physical effects (``$Corr$'' in Eq.~(\ref{eq_calc_LE})) that we need to apply to the {\tt GENIE} results.

First, we consider $\mu^-$ capture.  Because of the long muon lifetime relative to the short energy-loss time, both $\mu^-$ and $\mu^+$ come to rest in matter.  After that, $\mu^+$ decay, while $\mu^-$ undergo atomic capture, mostly on oxygen~\cite{Ponomarev:1973ya, Measday:2001yr, Czarnecki:2014cxa}. Then, our simulations using FLUKA~\cite{Ferrari:2005zk} show that $\simeq 21\%$ of $\mu^-$ undergo nuclear capture, producing a $\simeq 100$ MeV neutrino, which does not interact in the detector.  Although $\mu^-$ nuclear captures may also eject nucleons and lead to nuclear de-excitation gamma rays, we find that they cannot mimic DSNB signal events in the energy range $E_e =$ 16--90 MeV.  To account for $\mu^-$ capture, we reduce the overall yield of $\mu^-$ by $\simeq 21\%$, which affects only the $\nu_\mu$ CC events.

Second, there is a contribution from invisible $\pi^+$ from neutrino NC interactions, which is missing in our treatment of ($\nu_\mu + \bar{\nu}_\mu$) CC events as the source of invisible muons~\cite{Super-Kamiokande:2021jaq, Bays:2012wty}. A $\pi^+$ below the Cherenkov threshold will decay to an invisible muon. Therefore, if a neutrino interaction produces an invisible $\pi^+$ and no other visible particles, it could mimic the DSNB signal. Such events can come from all-flavor NC interactions with oxygen or hydrogen, which we refer to as the NC$ \pi^+$ channel (as noted above, $\pi^-$ undergo nuclear capture).  There are negligible corrections from $\pi^+$ interacting before decaying and from $\pi^-$ decaying in flight. We call the calculations to this point our ``Standard'' calculation. 

Third, Coulomb corrections due to outgoing charged particles being affected by the nuclear field should be included. The distortion effects increase for low charged-particle energies or high nuclear charges, and are thus important for this and the following sections.  For negatively charged particles, the Coulomb attraction lowers their outgoing momentum while increasing their production amplitude; the opposite occurs for positively charged particles. Our paper is the first to include the Coulomb corrections to this problem.

Widely used Coulomb correction methods are the Fermi function and the effective momentum approximation (EMA)~\cite{Traini:1988hs, Giusti:1987eoi}.  The former only works well for electrons below $\sim 10$~MeV, and the latter only for scattering of ultrarelativistic electrons on nuclei.  Here we use the modified EMA method~\cite{Engel:1997fy}, which works well for muons and electrons at the energies relevant to us. (It should also work well for charged pions.) In this method, the nucleus is approximated as a uniformly charged sphere with radius $R$ and the neutrino is assumed to interact at its center. The outgoing charged particles are thus initially subject to an electrostatic potential of $V = \pm3Z\alpha/2R$, which is about 5.3 MeV for oxygen and 1.7 MeV for hydrogen. This potential induces a shift in the total energy of the charged particle, i.e., 
\begin{equation}
    E \rightarrow E^{\rm eff} = E + V \,,
\end{equation}
and a rescaling of the scattering amplitude, i.e.,
\begin{equation}
\mathcal{M} \rightarrow \mathcal{M} \sqrt{\frac{k_{\rm eff} E_{\rm eff}}{k E}} \,,
\end{equation}
where $k$ and $k_{\rm eff}$ are the momentum of the charged particle before and after the shift, respectively.  These changes are in the same sense because the invisible muons have a spectrum that rises with increasing momentum.

For the LFG-NAV model set, as the Coulomb corrections are already included in the CCQES model, we apply these corrections only to the NC $\pi^+$ channels.  For the RFG-LS model set, we apply Coulomb corrections to all CC and NC $\pi^+$ channels.  For the CC channels, the correction increases the number of invisible muons in the $\nu_\mu$+O CC channel and decreases those in the $\bar{\nu}_\mu$+O CC, $\bar{\nu}_\mu$+H CC, and NC $\pi^+$ channels.  For the NC $\pi^+$ channel, which comes mostly from the decay of $\Delta$ baryons inside the nucleus, we ignore any corrections to the $\Delta$ baryons due to their large mass and short lifetime.  For the $\pi^+$ from $\Delta$ decay, we do not apply the correction to the scattering amplitude because they are not directly produced by neutrino interactions, but we do include the energy shift.  We call the calculations to this point our ``Coulomb'' calculation.

\begin{figure}[t]
\includegraphics[width=\columnwidth]{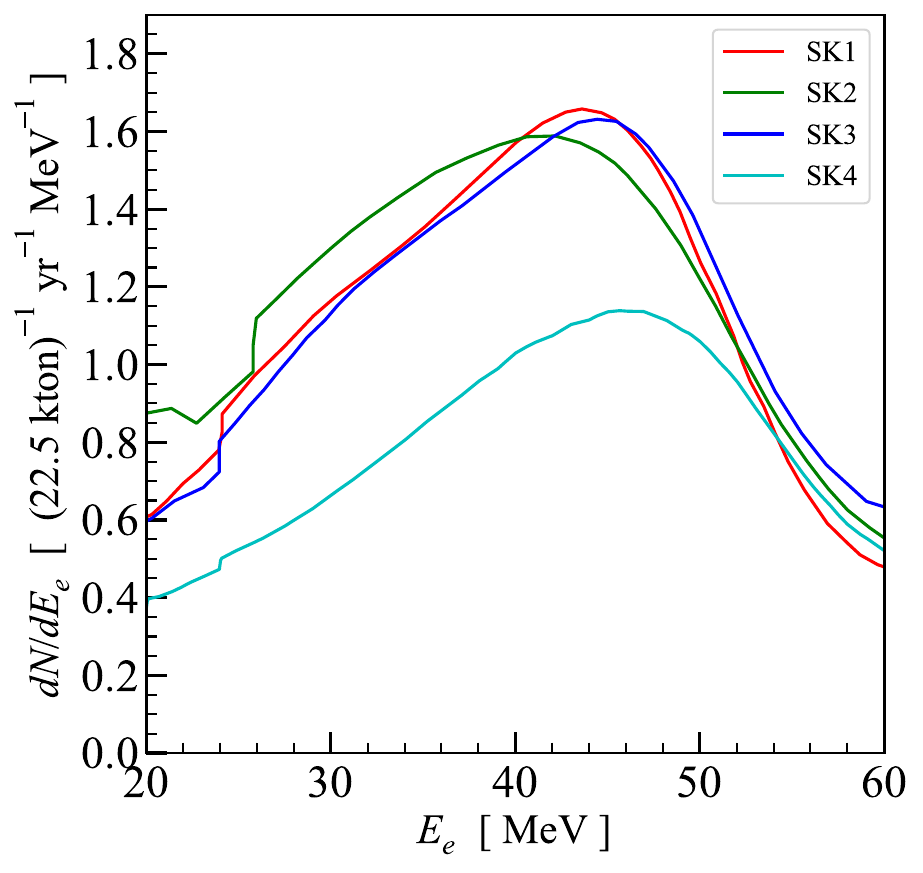}
\caption{A test of the consistency of Super-K's analysis results across all four Super-K stages.  The integrated results for stage IV are significantly lower than the average of stages I--III, for which the reason is unknown. Note that the difference is not due to an overall factor; the ratio is energy dependent.}
\label{fig_SK_phases_compare}
\end{figure}

\subsubsection{Detection effects}
\label{sec_repLEinvmu_calc_threshold}

The energy resolution and efficiency differ somewhat over Super-K stages I--IV, and we take this into account.  While the energy resolution is somewhat worse in stage II due to the reduced number of PMTs, it remains good enough that it barely changes the results.  The raw detection efficiency is near unity in all phases, though the analysis (or signal) efficiency is reduced by various cuts.   The analysis efficiency is $\simeq 90\%$ for most energies in Super-K stages I--III, while it is $\simeq 70$--85\% in stage IV~\cite{super_kamiokande_collaboration_2021_5779986}.  This is likely due to a new cut introduced for stage IV only~\cite{MichaelSmy}.  This cut, which is on the average charge deposit per PMT (see Sec.~VC5 and Fig.~14 in Ref.~\cite{Super-Kamiokande:2021jaq}), reduces the efficiency of detecting higher-energy events.

Figure~\ref{fig_SK_phases_compare} shows a consistency test of the four Super-K stages, focusing on the bump data~\cite{Super-Kamiokande:2021jaq}, where we use the Super-K fits instead of the data for visual clarity.  We correct for the different livetimes and efficiencies of each stage (including the new cut in stage IV), so that the curves for the four stages should agree.  For stage IV, the integrated spectrum is lower than the average of stages I--III by $\simeq35\%$, which has a statistical significance of $\simeq 7\sigma$.  The reason for this discrepancy is unknown, but we must bear it in mind when comparing to data.  We emphasize that we have made extensive efforts to find the source of this discrepancy, including through consulting many Super-K collaborators, but we have been unsuccessful.  A detailed study by Super-K is needed.

In the calculations above, we used the theoretical Cherenkov threshold, $\beta_{th} = 1/n$, to assess if charged particles should be cut. However, if a charged particle does not have enough energy above the threshold, it will not actually produce enough Cherenkov photons to trigger the detector. Therefore, a practical Cherenkov threshold should be calculated.  This is important because the spectra of charged particles in this energy range are rising with increasing momentum. Therefore, for the first time, we tackle this problem.

For Super-K, barely relativistic muons can be identified~\cite{Super-Kamiokande:2011lwo}.  This works as long as the muon itself activates at least the super-low energy (SLE) trigger. This trigger has a threshold of 17--24 photoelectrons, where the precise value depends on the analysis period. We use the lowest number, 17, as it is believed that even a lower number of photoelectrons may still have a chance to be identified~\cite{MichaelSmy}.  We convert this number of photoelectrons to a corrected threshold for $p_\mu$.  
First, we assume the number of photoelectrons equals the number of PMTs hit, as the total number of PMTs in Super-K's inner detector, $> 10^4$, is much larger.  
Considering the coverage and quantum efficiency of Super-K PMTs, this corresponds to about 340 Cherenkov photons with wavelengths $\lambda \simeq 300$--700 nm emitted by the muon.  
Finally, we convert the number of Cherenkov photons, $N_{\rm ph}$, to the momentum of the charged particles using~\cite{Beacom:2003zu, ParticleDataGroup:2022pth}
\begin{equation}
\frac{d^2 N_{\rm ph}}{d x d \lambda}=\frac{2 \pi \alpha}{\lambda^2} \left[1-\frac{1}{\beta(x)^2 n^2(\lambda)}\right] \,,
\end{equation}
where $x$ is the distance that a charged particle travels, $\alpha$ the fine structure constant, $\beta(x)$ the velocity of the particle, and $n(\lambda) \simeq 1.33$ is the refractive index of water.  For muons, this changes the Cherenkov threshold from $p_\mu \simeq 120$~MeV to $p_\mu \simeq 135$~MeV.  For charged pions, this changes the Cherenkov threshold from 159~MeV to 176 MeV. Our calculation for the revised thresholds gives results consistent with those from practical Super-K detector simulations~\cite{ChenyuanXu}.  These changes lead to increased numbers of predicted invisible-muon events.


\subsection{Summary of predictions and uncertainties}
\label{sec_repLEinvmu_summary}

Table~\ref{tab_invmu_SK1234} summarizes our predicted numbers of decay electrons from invisible muons for Super-K stages I--IV for the two model sets of {\tt GENIE} (LFG-NAV and RFG-LS; see Sec.~\ref{sec_morebkgd_nuInt}). For LFG-NAV, the predictions match the measurements of Super-K stages I--III quite well and are somewhat high for Super-K stage IV, likely due to the efficiency issue noted above (Fig.~\ref{fig_SK_phases_compare} and related text).  For RFG-LS, the predictions are always higher than the measurements, especially for Super-K stage IV. 

\begin{figure*}[ht!!]
\includegraphics[width=\columnwidth]{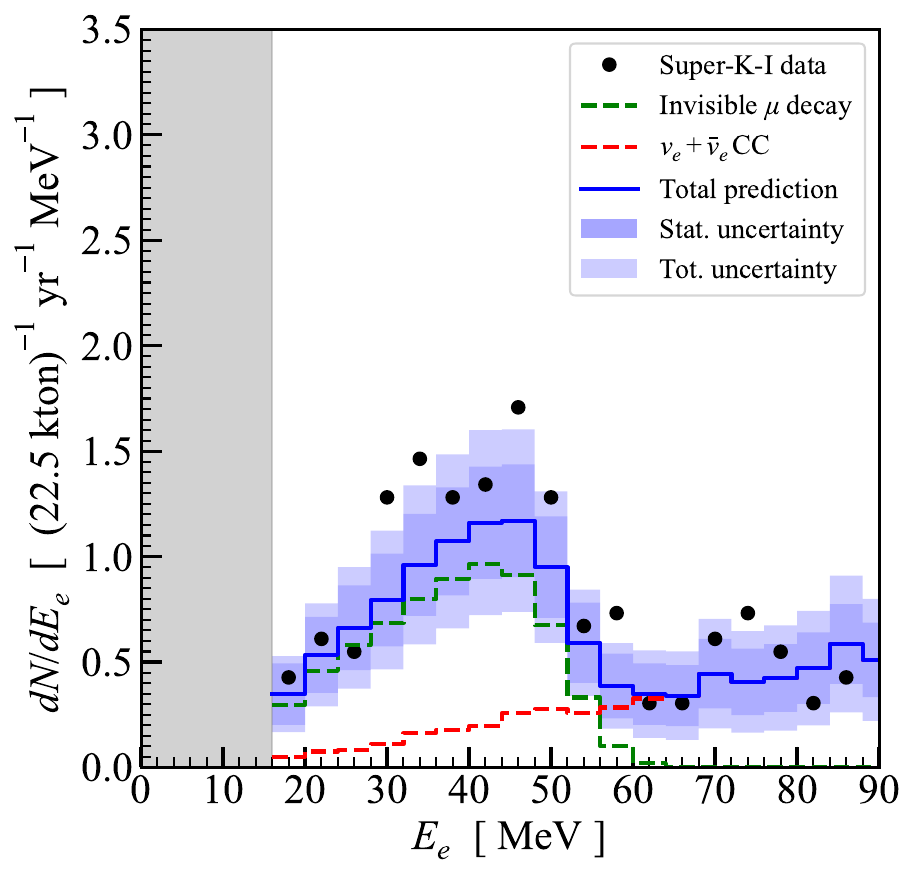}
\includegraphics[width=\columnwidth]{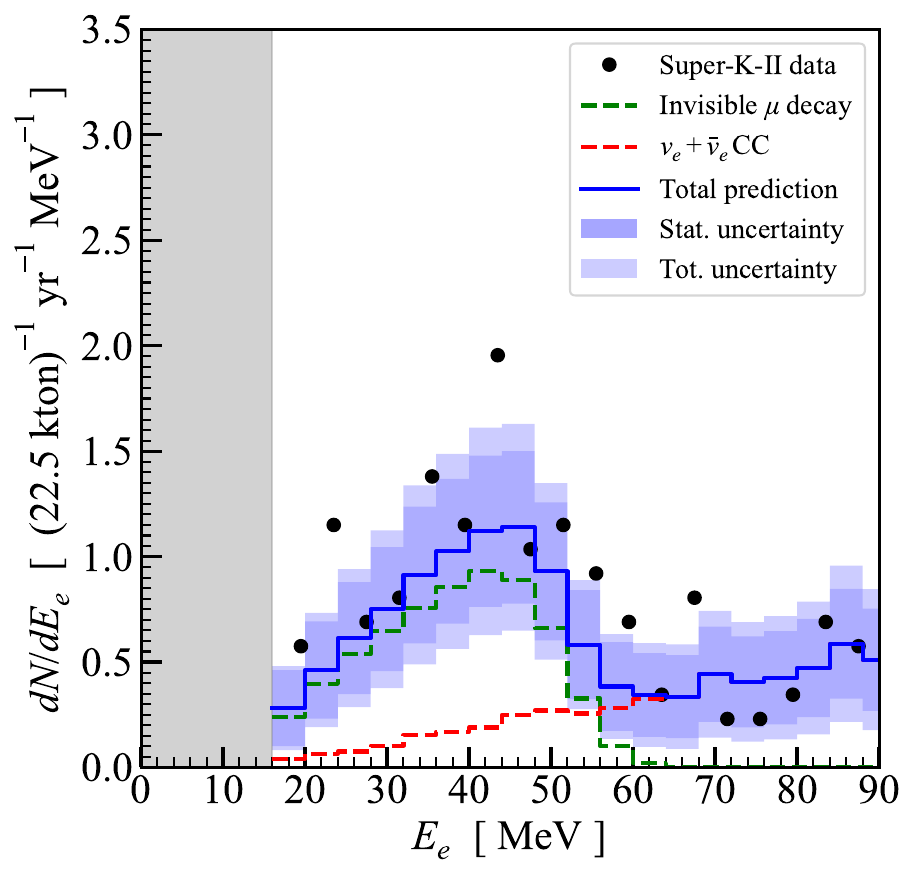}
\includegraphics[width=\columnwidth]{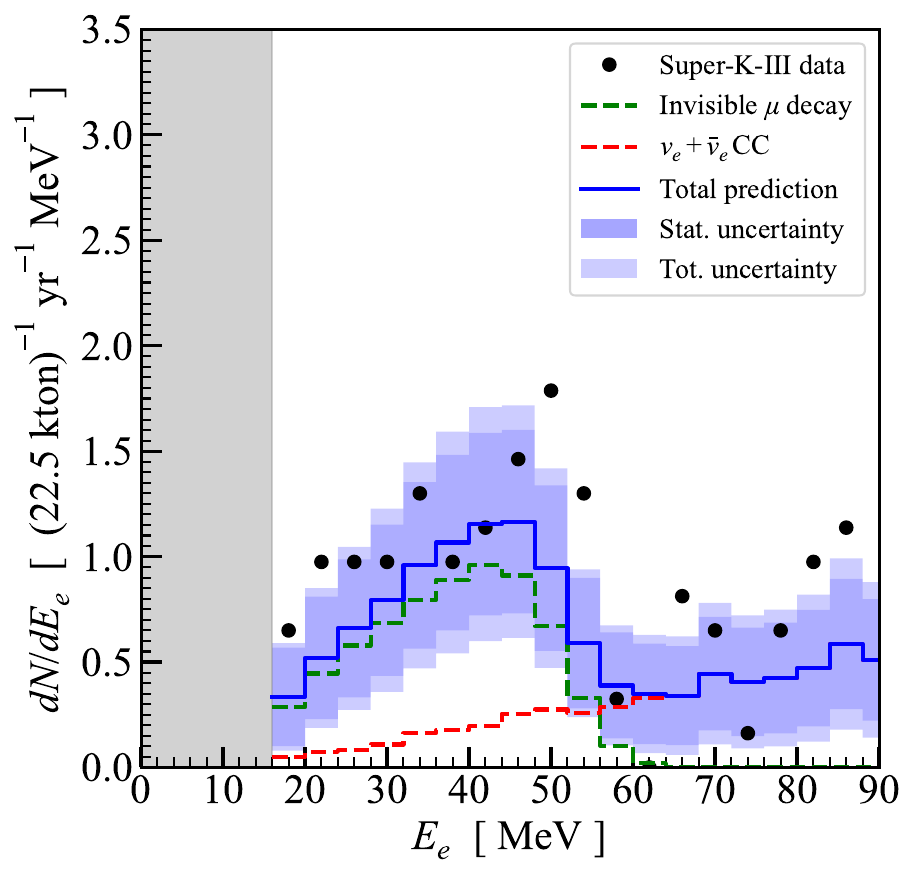}
\includegraphics[width=\columnwidth]{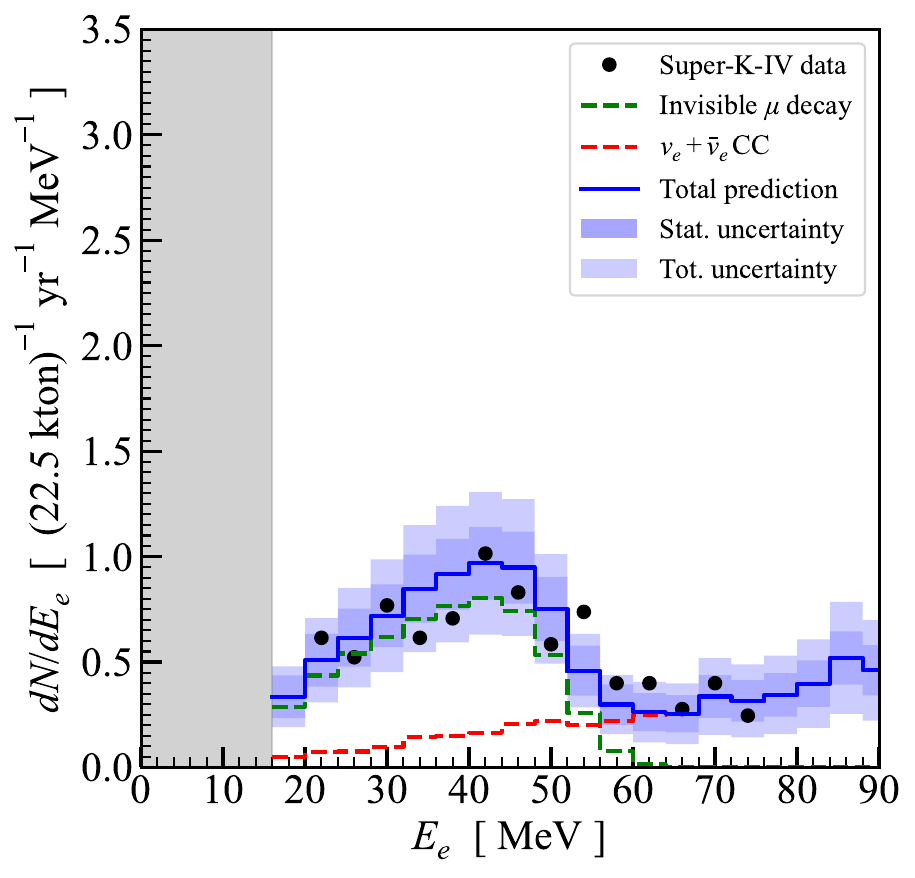}
\caption{Our complete calculations of the atmospheric-neutrino backgrounds (using the LFG-NAV model set) compared to all four stages of Super-K data~\cite{Super-Kamiokande:2021jaq}.  For the predictions, we plot the components in dashed lines and the totals in blue solid lines, with the uncertainties shown with blue bands.  Although we show 4-MeV steps, we have converted the values to units of (1 MeV)$^{-1}$ (i.e., a step of height 1 contains four counts).  Our calculations match the Super-K data reasonably well.
\vspace{0.5cm}
}
\label{fig_reproduce_LE_LFG}
\end{figure*}

Figure~\ref{fig_reproduce_LE_LFG} compares our LFG-NAV predictions to Super-K data in stages I--IV.  (As above, we show the uncertainties on the models instead of the data.)  For the invisible-muon component (the bump), we predict the normalization and the shape is known.  For LFG-NAV (Fig.~\ref{fig_reproduce_LE_LFG}), the agreement is reasonable for all four stages, and it would not have been so without the detailed calculations above. 
For the RFG-LS model set (see Fig.~\ref{fig_reproduce_LE_RFG} in the Appendix), the prediction is overall higher than the RFG-LS model set for all four stages and higher than the data for stage IV (see also Fig.~\ref{fig_SK_phases_compare}). 
This is expected as its MEC component is known to be too large~\cite{Dytman}, which explains why the LFG-NAV and RFG-LS models differ by more than the nominal $\simeq20\%$ cross section uncertainties (Sec.~\ref{sec_morebkgd_nuInt}).

Table~\ref{tab_invmu_SK4} shows our detailed predictions in Super-K stage IV for both the LFG-NAV and RFG-LS model sets.  The results for other stages (not shown) are in similar proportions.  The LFG-NAV model set embeds the best theoretical modeling elements implemented in {\tt GENIE} so far~\cite{GENIE3_manual}.  For each model set, we show the results for the major calculational steps, represented by different columns from left to right. Until new cuts based on secondary particles are developed~\cite{DSNB2}, all of the interaction channels that contribute to a given background component are indistinguishable.
\begin{itemize}

\item
``Naive" case (Sec.~\ref{sec_repLEinvmu_calc_eventClasses}).  This is a zeroth-order approximation, with the yields of the four CC channels following their corresponding fluxes and cross sections, as in Figs.~\ref{fig_atm_flux} and ~\ref{fig_xsec}.

\item
``Standard" case (Sec.~\ref{sec_repLEinvmu_calc_PhysCorr}).  We cut events with pion production and $\mu^-$ capture. For channels with oxygen targets, which are dominant, this removes large fractions of their events. For $\nu_\mu$+H CC, all events are removed due to there being no CCQES events. For $\bar{\nu}_\mu$+H CC, all events survive because CCQES dominates.  We add NC $\pi^+$ channels, which contribute about 30\% in the LFG-NAV model set and 20\% in the RFG-LS model set, an important point that was not noted before.

\item
``Coulomb" case (Sec.~\ref{sec_repLEinvmu_calc_PhysCorr}).  
For the LFG-NAV model set, the Coulomb correction is already included in the CC channels so we apply it to the NC $\pi^+$ channels only. For the RFG-LS model set, we apply it to all channels.  Coulomb corrections increase the yield of the $\nu_\mu$+O CC channel by $\simeq 35\%$ and decrease the yields of the $\bar{\nu}_\mu$+O CC, $\bar{\nu}_\mu$+H CC, and NC $\pi^+$ channels by $\simeq 25\%$, 10\%, and 10\%.

\item
``Threshold" case (Sec.~\ref{sec_repLEinvmu_calc_threshold}).  We correct the Cherenkov threshold from its theoretical value because barely relativistic charged particles may not trigger the detector.  This increases the yields by $\simeq 30\%$.

\item 
Nuclear gamma rays (Sec.~\ref{sec_repLEinvmu_calc_eventClasses}).  We show two cases for the efficiency of Super-K cuts on nuclear gamma rays, as this is uncertain.  We consider that either none have been removed ($\epsilon_\gamma = 0\%$) or that all of them have ($\epsilon_\gamma = 100\%$).  To decide between these, further information from Super-K is needed.

\end{itemize}

Table~\ref{tab_invmu_unc} summarizes our estimated systematic uncertainties in the predicted yields.
For the atmospheric-neutrino fluxes, we estimate the uncertainties to be $\simeq 20\%$ (Sec.~\ref{sec_morebkgd_atmNu}).  For the neutrino-nucleus cross sections, we also estimate the uncertainties to be $\simeq 20\%$ (Sec.~\ref{sec_morebkgd_nuInt}).  For the Cherenkov threshold correction, due to a lack of information, we estimate an uncertainty of $\simeq 20\%$.  Adding these uncertainties in quadrature gives a total of $\simeq 35\%$. This is large enough to cover other uncertainties that we do not specifically note.

For the nuclear gamma-ray branching ratio ($Br_\gamma$) and cut efficiency ($\epsilon_\gamma$), though they both have large uncertainties, comparing our final calculation with Super-K data gives important insights.   The consistency between data and LFG-NAV prediction indicates that the choice of $Br_\gamma \times \epsilon_\gamma \simeq 50\%\times100\%$ that we use is good.  For the RFG-LS model set (Fig.~\ref{fig_reproduce_LE_RFG} in Appendix~\ref{app_RFG}), lowering $\epsilon_\gamma$ would make the deviation between the prediction and the data even larger. Because we do not expect the true $Br_\gamma$ to be very different from $50\%$, this means that Super-K has likely already efficiently cut the invisible-muon background events with nuclear gamma rays, though this is not explicitly discussed in their papers~\cite{Malek:2002ns, Super-Kamiokande:2011lwo, Super-Kamiokande:2013ufi, Super-Kamiokande:2021jaq}.

\begin{table*}[t]
\caption{Summary of the total measured (using Super-K fits~\cite{Super-Kamiokande:2021jaq}) and predicted numbers of decay electrons ($E_e = $16--90~MeV) from invisible muons for Super-K stages I--IV, assuming $\epsilon_\gamma=100\%$.  The systematic uncertainty in our prediction is about 35\%.
The ratios are shown in parentheses.  For both model sets, the agreement is reasonable; the RFG-LS model set has some disagreement in stage IV, possibly due to the efficiency issue noted in the text.}
\label{tab_invmu_SK1234}
\smallskip
\renewcommand{\arraystretch}{1.1} \centering 
\vspace{0.25cm}
\begin{tabular}{c|c|c|c|c}

\hline 
\hhline{-----}
   & Stage I & Stage II & Stage III & Stage IV \\ 
\hline \hline

Measurement	&	146	&	74	&	50	&	155	\\
\hline									
LFG-NAV prediction (ratio to measurement)	&	106 (0.73)	&	55 (0.74)	&	40 (0.8)	&	185 (1.19)	\\
\hline									
RFG-LS prediction (ratio to measurement)	&	148 (1.01)	&	77 (1.04)	&	57 (1.14)	&	259 (1.67)	\\
\hline
\end{tabular}
\vspace{0.5cm}
\end{table*}

\begin{table*}[t]
\caption{
Detailed predicted numbers of decay electrons ($E_e =$ 16--90~MeV) from invisible muons in Super-K stage IV. \textbf{Top panel:} Results for the {\tt GENIE} LFG-NAV model set. \textbf{Bottom panel:} Results for the {\tt GENIE} RFG-LS model set.  Numbers in boldface are our final predictions.  The columns show the steps of including various effects, concluding with the ``Threshold" column.  The CC channels of the LFG-NAV model set (upper table) intrinsically include Coulomb corrections, so the numbers remain unchanged when moving from the ``Standard'' column.  We show results for two assumptions about nuclear gamma rays, where the discussion in the text favors $\epsilon_\gamma = 100\%$.  Note that while $Br_\gamma \simeq 50\%$, it varies between interaction channels. The total number of decay electrons passing all cuts in Super-K stage IV is 155 (from Super-K's fits to the data)~\cite{Super-Kamiokande:2021jaq}.
}
\label{tab_invmu_SK4}
\smallskip
\renewcommand{\arraystretch}{1.1} \centering
\vspace{0.25cm}
\begin{tabular}{c||c|c|c|c||c|c|c}
\hline 
    \multirow{2}{*}{Interaction channel}  & \multicolumn{4}{|c||}{$Br_\gamma=50\%$, $\epsilon_\gamma=0\%$} & \multicolumn{3}{c}{$Br_\gamma=50\%$, $\epsilon_\gamma=100\%$}  \\ 
\hhline{~-------}
   & Naive   & Standard & Coulomb & Threshold & Standard & Coulomb & Threshold \\ 
\hline \hline

$\nu_\mu$+O CC	&	159	&	107	&	107	&	143	&	56	&	56	&	75	\\
$\bar{\nu}_\mu$+O CC	&	35	&	30	&	30	&	39	&	14	&	14	&	19	\\
$\nu_\mu$+H CC	&	7	&	0	&	0	&	0	&	0	&	0	&	0	\\
$\bar{\nu}_\mu$+H CC	&	24	&	23	&	23	&	30	&	23	&	23	&	30	\\ \hline
NC $\pi^+$	&		&	92	&	84	&	107	&	51	&	46	&	61	\\ \hline
Total	&	226	&	253	&	245	&	319	&	145	&	140	&	{\bf 185 }	\\
Total/Super-K-IV (155)	&	1.45	&	1.63	&	1.58	&	2.05	&	0.93	&	0.90	&	{\bf 1.19 }	\\

\hline
\end{tabular}

\renewcommand{\arraystretch}{1.1} \centering 
\vspace{0.5cm}

\begin{tabular}{c||c|c|c|c||c|c|c}
\hline 
    \multirow{2}{*}{Interaction channel}  & \multicolumn{4}{|c||}{$Br_\gamma=50\%$, $\epsilon_\gamma=0\%$} & \multicolumn{3}{c}{$Br_\gamma=50\%$, $\epsilon_\gamma=100\%$}  \\ 
\hhline{~-------}
   & Naive   & Standard & Coulomb & Threshold & Standard & Coulomb & Threshold \\ 
\hline \hline

$\nu_\mu$+O CC	&	223	&	158	&	215	&	282	&	84	&	115	&	150	\\
$\bar{\nu}_\mu$+O CC	&	48	&	40	&	29	&	42	&	20	&	15	&	20	\\
$\nu_\mu$+H CC	&	9	&	0	&	0	&	0	&	0	&	0	&	0	\\
$\bar{\nu}_\mu$+H CC	&	24	&	23	&	21	&	27	&	23	&	21	&	27	\\ \hline
NC $\pi^+$	&		&	96	&	88	&	111	&	52	&	48	&	61	\\ \hline
Total	&	302	&	317	&	353	&	461	&	178	&	198	&	{\bf 259 }	\\
Total/Super-K-IV (155)	&	1.94	&	2.04	&	2.27	&	2.96	&	1.15	&	1.27	&	{\bf 1.67 }	\\

\hline 
\end{tabular}
\end{table*}

\begin{table*}[t]
\caption{Summary of the systematic uncertainties in our calculation of the decay electrons.}
\label{tab_invmu_unc}
\medskip
\renewcommand{\arraystretch}{1.1} \centering 
\begin{tabular}{c|c|c|c||c}
\hline 
\hhline{-----}
   & Atmospheric neutrino fluxes & Neutrino interactions & Cherenkov threshold & \ \ Total\ \ \  \\ 
\hline \hline

Uncertainty	&	20\%	&	20\%	&	20\%	& 35\% \\
\hline									
\end{tabular}
\vspace{0.5cm}
\end{table*}


\section{New results on Super-K Low-Energy Atmospheric data: \\ ($\nu_e + \bar{\nu}_e$) CC component}
\label{sec_repLEnuecc}

In this section, we calculate our predictions for the ($\nu_e + \bar{\nu}_e$) CC component (the ramp) of the atmospheric-neutrino background, the smaller of the two components.  Previously, in Super-K's Refs.~\cite{Malek:2002ns, Super-Kamiokande:2011lwo, Super-Kamiokande:2013ufi, Super-Kamiokande:2021jaq}, neither the shape nor the normalization was given, though their Ref.~\cite{Super-Kamiokande:2021jaq} mentions that modeling was done.  These interactions produce a single primary electron.  While these events have some directionality in principle, the low statistics and the near-isotropy of the atmospheric-neutrino flux make this hard to exploit.

Our calculation is similar to that for the invisible muon component, but there are some differences.  First, here we must also take into account the atmospheric-neutrino fluxes below 100 MeV, for which we use the results of Ref.~\cite{Battistoni:2005pd}.  Second, here we assume that, effectively, $\epsilon_\gamma = 0\%$ because the energy in any nuclear gamma rays (and the secondary electrons they produce) is much less than that in the primary electron from the CC interaction.  Third, here no NC channels contribute.

The uncertainty in calculating the ($\nu_e + \bar{\nu}_e$) CC component is larger than for the invisible muon component.  The neutrino fluxes are more uncertain because the energies are lower and because the $\nu_e$ and $\bar{\nu}_e$ fluxes arise from a further step in the decay chains, i.e., $\pi \rightarrow \mu \rightarrow e$.  The neutrino cross sections are more uncertain because the interactions are more in the nuclear than the particle regime.  Quantitatively, we expect the total uncertainty to be $\simeq$ 45\%, arising from uncertainties in the flux (25\%), cross section (30\%), and Coulomb correction (20\%), combined in quadrature.

Figure~\ref{fig_reproduce_LE_LFG} compares our LFG-NAV predictions of the ($\nu_e+\bar{\nu}_e$) CC component to the data in Super-K stages I--IV. The spectrum rises because both the flux (Fig.~\ref{fig_atm_flux}) and cross section (Fig.~\ref{fig_xsec}) do. With the detailed calculations described above, the predictions from LFG-NAV (Fig.~\ref{fig_reproduce_LE_LFG}) agree reasonably well with the data, though the uncertainties are large.  There is one peculiarity to mention.  In stages I--III, the data ranged up to 90 MeV.  In stage III, there are some rather high points at the highest energies.  Then, in stage IV, Super-K truncated the highest energies without explanation.  In fact, we recommend extending, not truncating, the range. Figure~\ref{fig_reproduce_LE_RFG} in the Appendix shows the same for the RFG-LS model set, where, as for the invisible muon component, the predictions are somewhat too large for stage IV  (see also Fig.~\ref{fig_SK_phases_compare}). Again, the results from the RFG-LS model set are overall higher than those from the LFG-NAV model set due to its larger MEC component~\cite{Dytman}.

\begin{figure*}[t]
\includegraphics[width=\columnwidth]{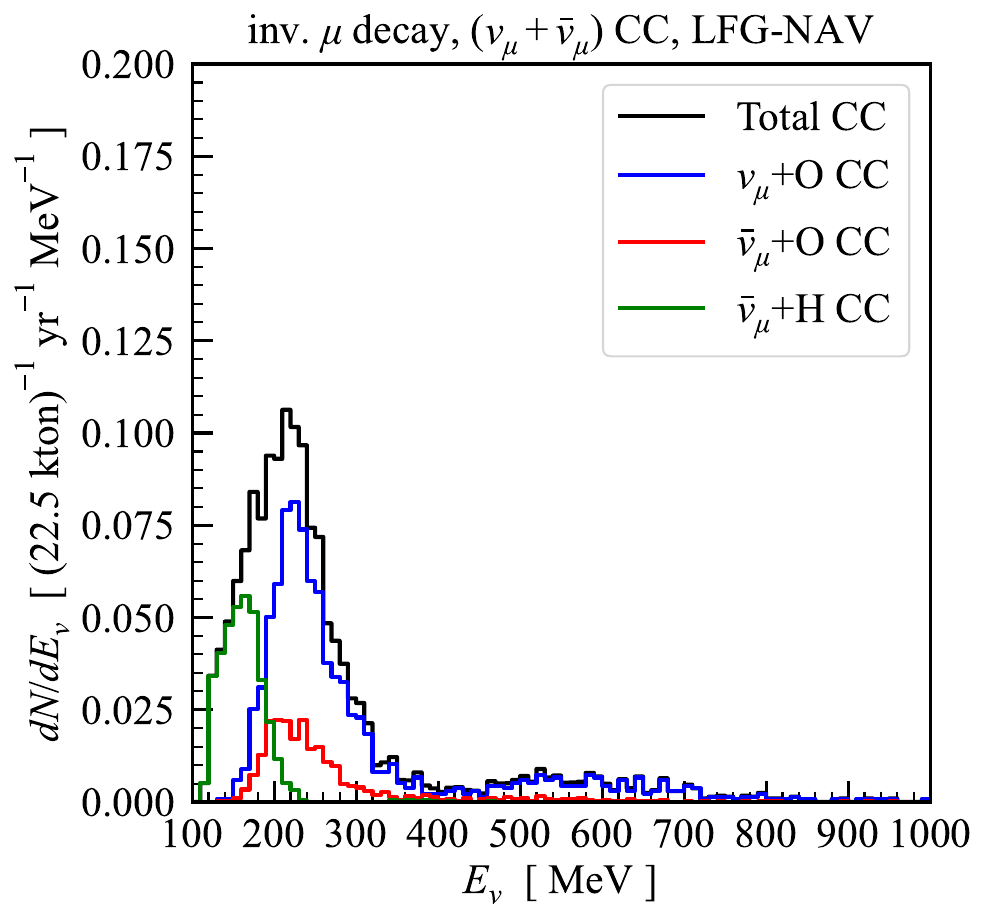}
\includegraphics[width=\columnwidth]{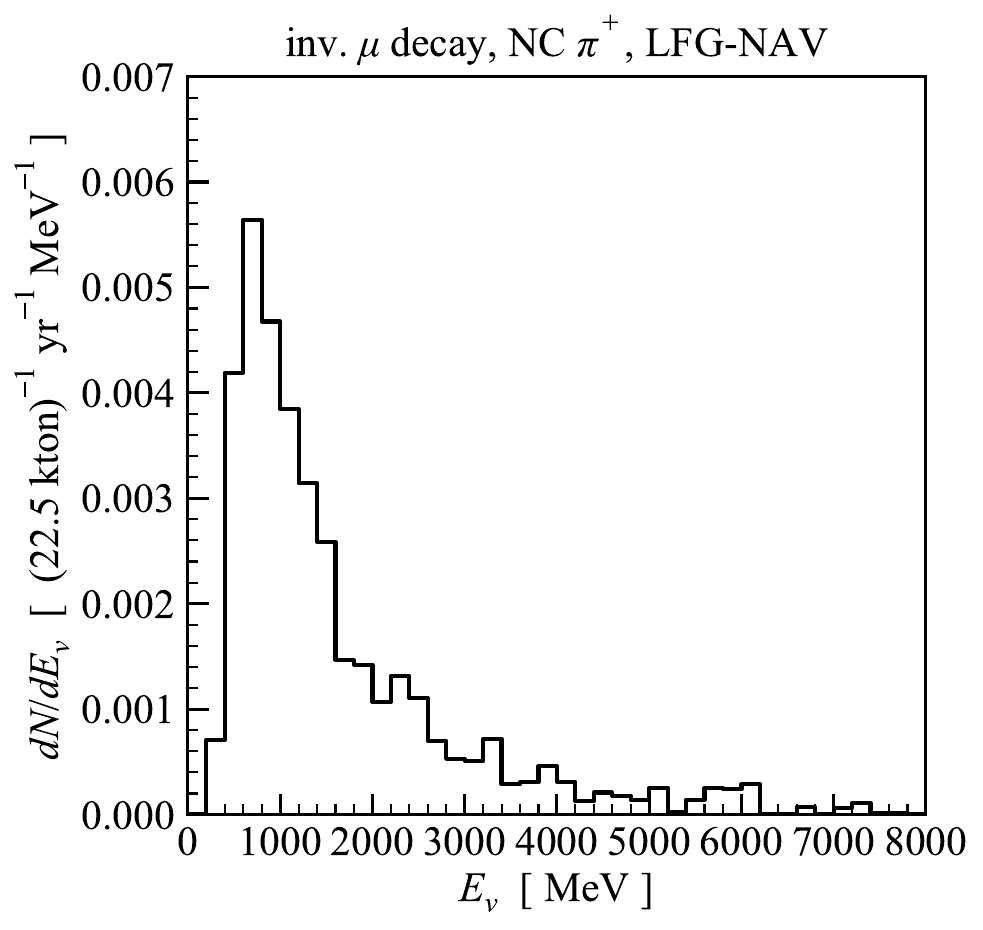}
\caption{Spectra of parent neutrinos for the invisible-muon component of the atmospheric-neutrino backgrounds in Super-K stage IV, calculated for the LFG-NAV model set.   {\bf Left:} ($\nu_\mu + \bar{\nu}_\mu$) CC component ($\simeq 70\%$ of total).  {\bf Right:} NC $\pi^+$ component ($\simeq 30\%$ of total); note changes in the axis ranges. For the RFG-LS model set, see Fig.~\ref{fig_parent_spec_invmu_RFG} in Appendix~\ref{app_RFG}.}
\label{fig_parent_spec_invmu_LFG}
\end{figure*}

\begin{figure*}[t]
\includegraphics[width=0.978\columnwidth]{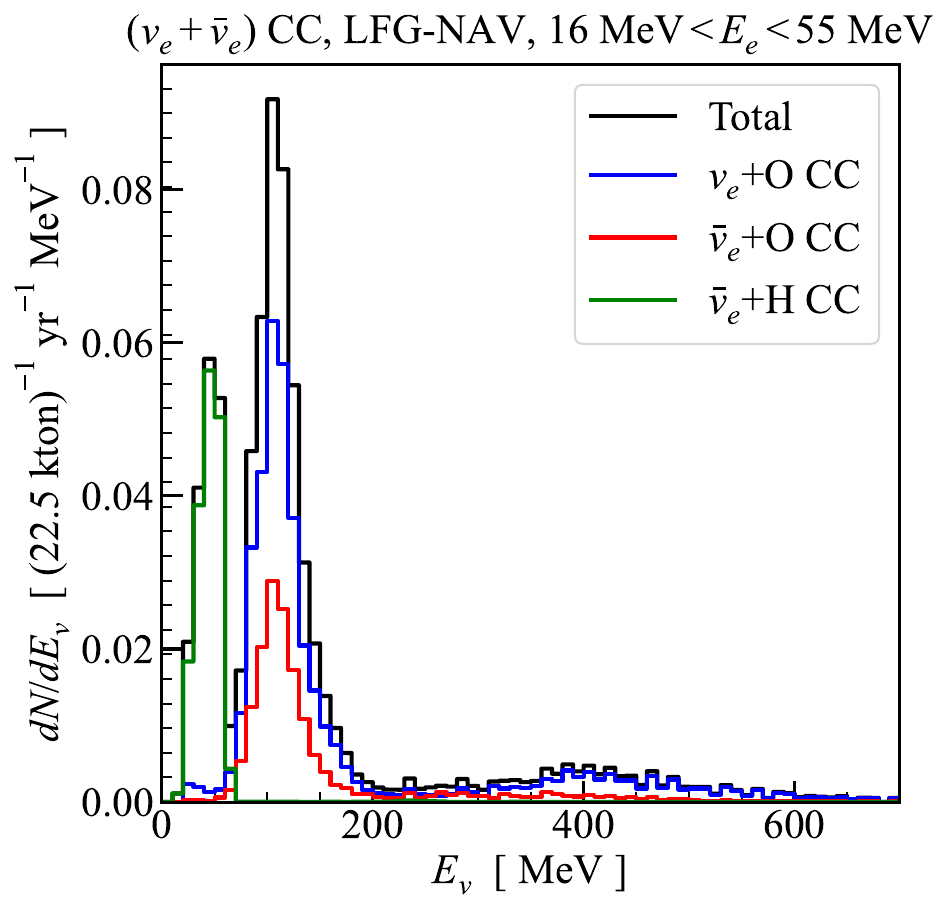}
\includegraphics[width=\columnwidth]{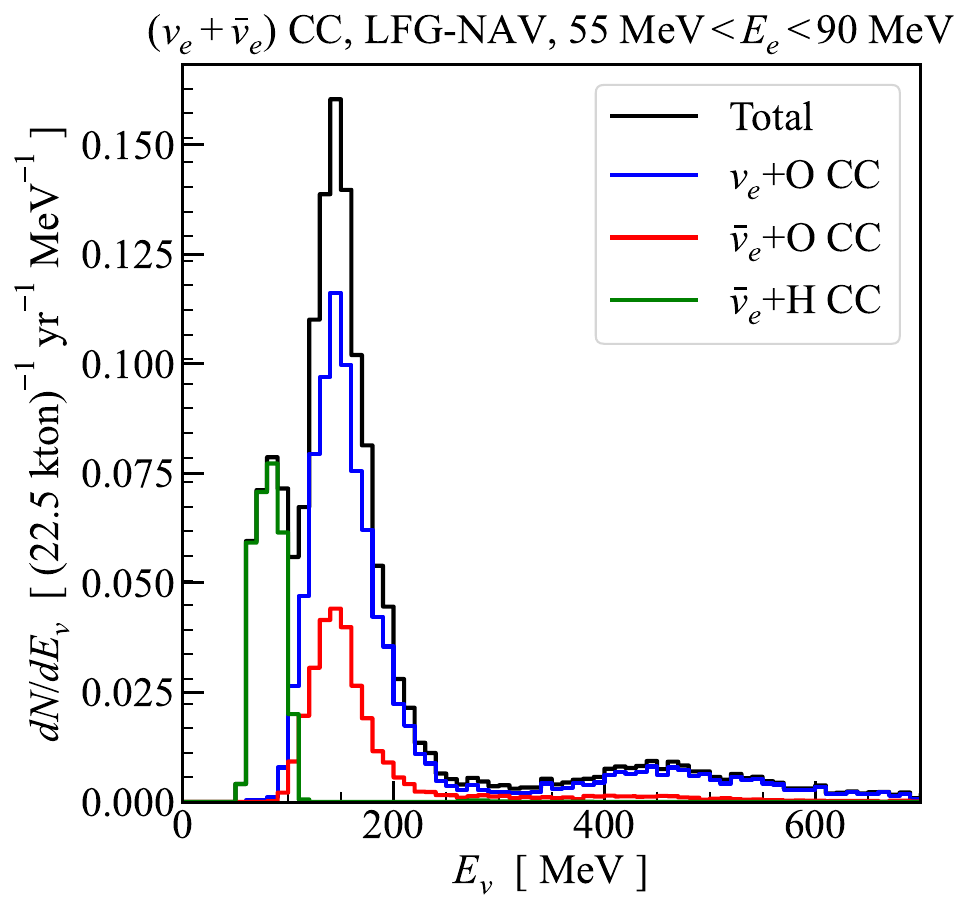}
\caption{Spectra of parent neutrinos for the ($\nu_e + \bar{\nu}_e$) CC component of the atmospheric-neutrino backgrounds in Super-K stage IV, calculated for the LFG-NAV model set (Sec.~\ref{sec_repLEnuecc}). {\bf Left:}  Results for $E_e =$ 16--55 MeV (45\% of the total counts). {\bf Right:}  Results for $E_e =$ 55--90 MeV (55\% of the total counts). For the RFG-LS model set, see Fig.~\ref{fig_parent_spec_nuecc_RFG} in Appendix~\ref{app_RFG}.}
\label{fig_parent_spec_nuecc_LFG}
\end{figure*}


\section{Parent Neutrino Spectra}
\label{sec_LEparent}

In this section, we calculate the parent neutrino spectra for the two components of the low-energy atmospheric-neutrino backgrounds.  These spectra are needed to determine how to best focus work on reducing uncertainties.

Figure~\ref{fig_parent_spec_invmu_LFG} shows our LFG-NAV results for the invisible-muon component for Super-K stage IV.  For the CC contributions (left panel), the dominant part is from $\nu_\mu + {\rm O}$ CC interactions, with a smaller part from $\bar{\nu}_\mu + {\rm O}$ CC interactions, both in the range $\simeq100$--400 MeV, mostly due to CCQES.  The small bump between $\simeq 400$ and 800 MeV is due to MEC and RES contributions.  There is also a contribution from $\bar{\nu}_\mu + {\rm H}$ CC interactions that starts at lower energies due to having no nuclear threshold.  In a simple CCQES model, the $\bar{\nu}_\mu + {\rm O}$ and $\bar{\nu}_\mu + {\rm H}$ interactions would produce neutrons that can be tagged by capture on gadolinium, while the $\nu_\mu + {\rm O}$ would not.  However, as we show in Ref.~\cite{DSNB2}, the more realistic picture is in fact more favorable, with a substantial fraction of $\nu_\mu + {\rm O}$ events also producing neutrons due to nuclear effects.  For the NC contribution (right panel), the dominant part is from $\Delta$ resonance production that decays to a sub-Cherenkov $\pi^+$.  The large majority of these NC events produce neutrons.

Figure~\ref{fig_parent_spec_nuecc_LFG} shows the results for the ($\nu_e + \bar{\nu}_e$) CC component.  Most of the parent neutrinos are in the range 20--200~MeV, and the $\nu_e + {\rm O}$ channel is dominant. There is a small bump between about 300 and 600 MeV, again due to MEC and RES.  The relative contributions from different channels are similar to the CC component of the invisible muons.  We split the results at $E_e = 55$ MeV, roughly where the invisible-muon and ($\nu_e + \bar{\nu}_e$) CC components of the background cross in Fig.~\ref{fig_reproduce_LE_LFG}.

Overall, the shapes of these spectra primarily follow from the competition between the rising charged-particle spectra at low energies and the higher probability of final states at high energies being subject to cuts.  For both background components, the most striking point is that the neutrino energies are much higher than the measured electron energies. (For the  signal, these are separated by only $\simeq$ 1.3 MeV.)  For the invisible-muon component, this is because the electrons are produced through muon decay at rest, which obscures the initial muon energy.  For the ($\nu_e + \bar{\nu}_e$) CC component, this is because the neutrino interaction rate only becomes appreciable at high enough neutrino energies; the low observed electron energies arise through the low-energy tail of the neutrino-oxygen differential cross section, plus neutrino-hydrogen interactions.

The fact that many of the background parent neutrino energies are relatively high is encouraging for cutting those events through the other secondary particles that are produced but for which cuts have yet to be devised.  For example, the DSNB-induced neutrons are mostly $< 1$~MeV, but atmospheric-neutrino-induced neutrons could be as energetic as 100~MeV and can travel much further~\cite{Beacom:2003nk}.  These points are detailed in our forthcoming paper~\cite{DSNB2}.  Knowing the parent-neutrino energy distributions is also helpful for assessing the effects of neutrino mixing and solar modulation.


\section{Outline of Ways Forward}
\label{sec_discussion}

In this section, we discuss ways to reduce uncertainties on the predictions of the DSNB detector backgrounds.  While our results above are adequate to start to guide strategies to reduce these backgrounds, improved precision would help.  Because detection of the DSNB in Super-K is so close, even modest improvements are important.  Here we focus on the most significant needs.


\subsection{Input data: atmospheric neutrino fluxes and neutrino-interaction modeling}

For the atmospheric-neutrino fluxes, the uncertainties are presently $\simeq 20\%$ in the neutrino energy ranges that generate most of the DSNB backgrounds, but these uncertainties should be reducible. The flux predictions that we use --- from Refs.~\cite{Honda:2015fha, Battistoni:2005pd} --- are based on old cosmic-ray data, whereas Ref.~\cite{Evans:2016obt} has shown that updating with more recent measurements, which have smaller uncertainties, can reduce the flux-prediction uncertainties by about a factor of two. New 3D calculations using these and other contemporary inputs are needed.

For the neutrino-nucleus cross sections, the uncertainties are presently at least $\simeq 20\%$ in the most important energy ranges, but these should also be reducible.  These cross sections are primarily characterized numerically through simulation codes~\cite{Andreopoulos:2009rq, Buss:2011mx, Golan:2012rfa, Isaacson:2022cwh, NEUT} that use a theoretical framework calibrated to neutrino and other scattering data, though direct neutrino-oxygen measurements~\cite{K2K:2006odf, T2K:2014vog, Super-Kamiokande:2019hga, T2K:2019zqh, T2K:2020jav, Sakai:2023xdi} are scarce.  These simulation codes should be updated to incorporate Coulomb and other corrections.  Finally, it would be valuable for the authors of the simulation codes to compare results on a variety of predictions and to use these results to better quantify uncertainties.

It is especially important to better characterize the emission of final-state particles besides charged leptons, as this will help develop better cuts.  As detailed in Sec.~\ref{sec_repLEinvmu_calc_GENIEchoices}, there are significant uncertainties about nuclear gamma-ray emission; these uncertainties need to be reduced through new experimental and theoretical studies.  Now that Super-K is running with added gadolinium, another critical question is about primary or secondary neutron emission, as explored in Ref.~\cite{DSNB2}.  The Sudbury Neutrino Observatory (SNO) (heavy water; completed) and SNO+ (light water phase; completed) experiments, while they have low rates of atmospheric-neutrino interactions, have the potential to make important measurements of gamma-ray and neutron production due to their very low background rates.  Measurements with ANNIE, an accelerator neutrino experiment with a gadolinium-loaded water target, will be important for measuring both cross sections and the emission of neutrons and gamma rays~\cite{ANNIE:2017nng, ANNIE:2019azm}.


\subsection{Super-K atmospheric analyses}

New work is urgently needed on Super-K analyses of atmospheric neutrinos.  (As a reminder, we focus on their pre-gadolinium data, as it has the longest exposure; first results from their DSNB search with post-gadolinium data are given in Ref.~\cite{Super-Kamiokande:2023xup}.)  One fundamental difficulty is that there are two disconnected data samples, split at roughly 100 MeV, and not quite overlapping. The higher-energy sample~\cite{Ashie:2005ik, Super-Kamiokande:2010orq, Super-Kamiokande:2011dgc, Super-Kamiokande:2015qek, Super-Kamiokande:2017yvm, Super-Kamiokande:2019gzr} is analyzed as a signal to measure neutrino mixing, but the lower-energy sample~\cite{Malek:2002ns, Super-Kamiokande:2011lwo, Super-Kamiokande:2013ufi, Super-Kamiokande:2021jaq} is treated only as a background to the DSNB.  For our purposes here, the deficiencies of the higher-energy analysis are that it uses only the cleanest events and does not adequately specify how neutrino events register in the detector.  The deficiencies of the lower-energy analysis are that it does not compare to theoretical predictions (hence the need for this paper) and also does not adequately specify how neutrino events register in the detector.  A new approach is needed to better connect these analyses, focusing on using all the data to accurately measure the fluxes and event rates, presenting results as a function of detected energy.  For this purpose, the neutrino-mixing parameters can be taken from laboratory experiments.

It would be very helpful for future Super-K DSNB papers to provide details comparable to what we do above.  In addition, key questions to resolve include:
\begin{enumerate}

\item For invisible-muon events \textit{with} nuclear gamma rays, what are the gamma-ray probabilities and energies?  For $(\nu_e + \bar{\nu}_e)$ CC interactions, can nuclear gamma rays be identified?

\item How do the spectra of the low-energy events ($<$100 MeV) in detected energy connect to those at energies up through a few hundred MeV?

\item What are detection thresholds for barely relativistic muons and pions (Sec.~\ref{sec_repLEinvmu_calc_threshold})?

\item Why are the low-energy spectra observed in Super-K stage IV inconsistent with those in earlier stages (Sec.~\ref{sec_repLEinvmu_calc_threshold})?

\item Thinking ahead to future analyses, what are the details of the spallation and atmospheric NC events below 16 MeV, both before and after cuts?

\end{enumerate}
Last, it would be helpful if Super-K would provide full event data for every low-energy event, as this would enable independent analyses.


\section{Conclusions}
\label{sec_conclusion}

The first detection of the DSNB will be of great importance, as it will test the neutrino emission per core collapse and the cosmic core-collapse rate.  Super-K is large enough to have collected $\sim 50$--100 DSNB events in total above an electron energy of $E_e = 16$ MeV (and many more at lower energies) in its $\gtrsim 25$~years of operation, but these events are presently obscured by detector backgrounds.  The largest backgrounds are due to atmospheric-neutrino interactions with nuclei.  The bump component arises from the electrons produced through the decays of invisible (sub-Cherenkov) muons, and the ramp component arises from electrons produced through ($\nu_e + \bar{\nu}_e$) CC interactions.

New theoretical work is needed to better understand the physical origins of these backgrounds and how to cut them further.  This matters both for reanalyzing past data as well as for making the most of new data since 2020, when Super-K began adding dissolved gadolinium to tag neutrons, which will greatly reduce detector backgrounds and allow a lower analysis threshold~\cite{Beacom:2003nk, Super-Kamiokande:2021the, Super-Kamiokande:2023xup}.

In this paper, we perform the first detailed calculations of the dominant atmospheric-neutrino backgrounds for DSNB searches in Super-K, taking into account neutrino mixing, neutrino-nucleus interactions, and how events register in Super-K.  As a bottom line, our calculations can reasonably reproduce Super-K's observed atmospheric-neutrino backgrounds in the range $E_e = 16$--90 MeV, which are mostly produced by neutrinos in the range up to about 400 MeV.  Our key results are shown in Fig.~\ref{fig_reproduce_LE_LFG}, Table~\ref{tab_invmu_SK1234}, and Table~\ref{tab_invmu_SK4}.  Achieving this agreement required taking into account several physical and detector effects, as well as checking that our calculations reasonably reproduce Super-K's GeV-range atmospheric-neutrino data.  The detailed results and comprehensive roadmap provided in this paper will help Super-K improve sensitivity to the DSNB.  In our next paper~\cite{DSNB2}, we go further by detailing proposed new cuts that take advantage of our new knowledge of how different processes contribute to the observed backgrounds.

This program of work will not only be useful for reducing backgrounds for DSNB (and dark matter~\cite{Palomares-Ruiz:2007trf, Bell:2020rkw, Bell:2022ycf}) searches.  Put another way, Super-K has a large atmospheric-neutrino dataset below about 100 MeV that has never been exploited as a signal.  The counts are large, about 50 events/year after cuts for about 25 years, so about 1250 events in total.  Without cuts, these event counts would be more than a factor of two larger.  Combined with data from other detectors, an exciting new frontier in low-energy atmospheric neutrinos could be opened~\cite{Super-Kamiokande:2019hga, Kelly:2019itm, Newstead:2020fie, Cheng:2020aaw, Cheng:2020oko, Chauhan:2021fzu, 
Kelly:2021jfs, Denton:2021rgt, Zhuang:2021rsg, Kelly:2023ugn, Suliga:2023pve, Sakai:2023xdi}.  This would allow new tests of neutrino mixing and neutrino-nucleus interactions.


\bigskip
\section*{Acknowledgments}

We are grateful for the helpful discussions with Yosuke Ashida, Matthew Dolan, Sonia El Hedri, Ed Kearns, Yusuke Koshio, Shirley Li, Stephan Meighen-Berger, Kenny Ng, Michael Smy, Anna Suliga, Mark Vagins, Michael Wagman, Linyan Wan, Thomas Wester, Wanwei Wu, Chenyuan Xu, and Guanying Zhu.  We are grateful to the {\tt GENIE} collaboration for providing a comprehensive framework for neutrino-nucleus interactions, and to Costas Andreopoulos, Steven Dytman, Robert Hatcher, and Gabriel Perdue for valuable help.  We also made use of {\tt FLUKA}, which simulates particle transport in matter.

For the later stages of this work, BZ was supported through his present position at Fermilab (managed by the Fermi Research Alliance, LLC, acting under Contract No.\ DE-AC02-07CH11359) and his recent past position at Johns Hopkins (with funding from the Simons Foundation), while JFB was supported by National Science Foundation Grant No.\ PHY-2310018.  For the early stages of this work, BZ was supported by National Science Foundation Grant No.\ PHY-1714479 and a Neutrino Theory Network Grant provided to Ohio State through Fermilab, while JFB was supported by National Science Foundation Grants No.\ PHY-2012955 and PHY-1714479.


\clearpage
\onecolumngrid

\appendix

\section{Results from the RFG-LS model set}
\label{app_RFG}

In this section, we show our predictions from the RFG-LS model set. The predictions are larger than Super-K data and the predictions for the LFG-NAV model set, which match Super-K data better. All the figures in this section are referred to in the main text.

\begin{figure*}[b!]
\includegraphics[width=0.49\textwidth]{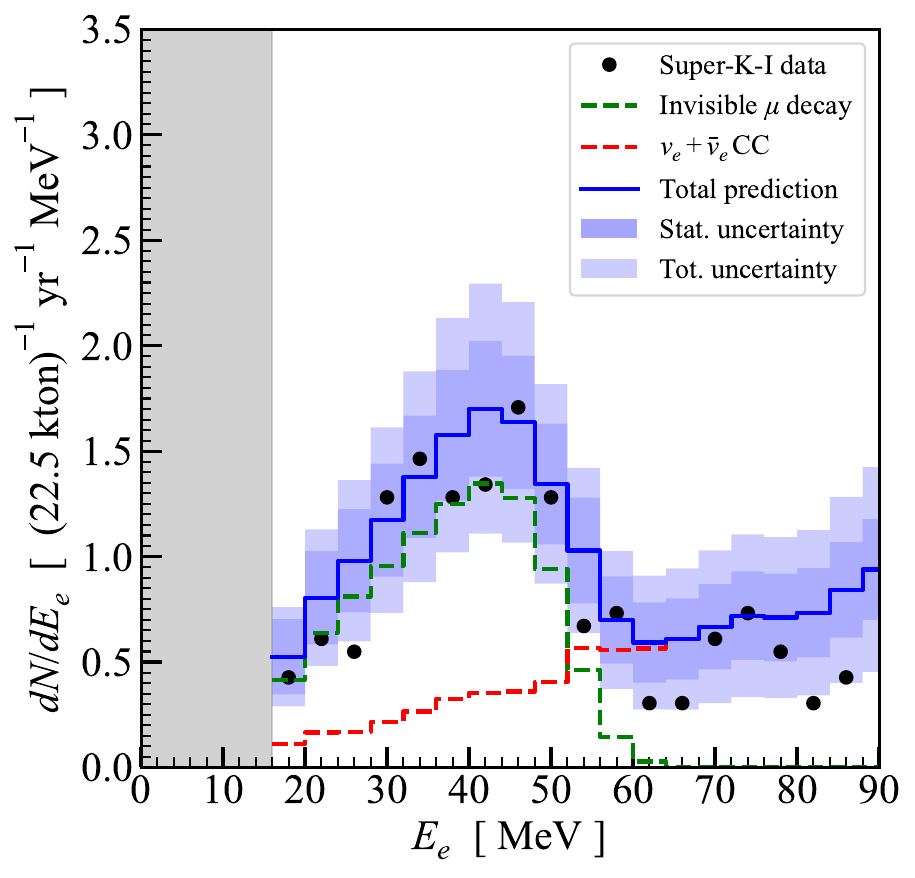}
\includegraphics[width=0.49\textwidth]{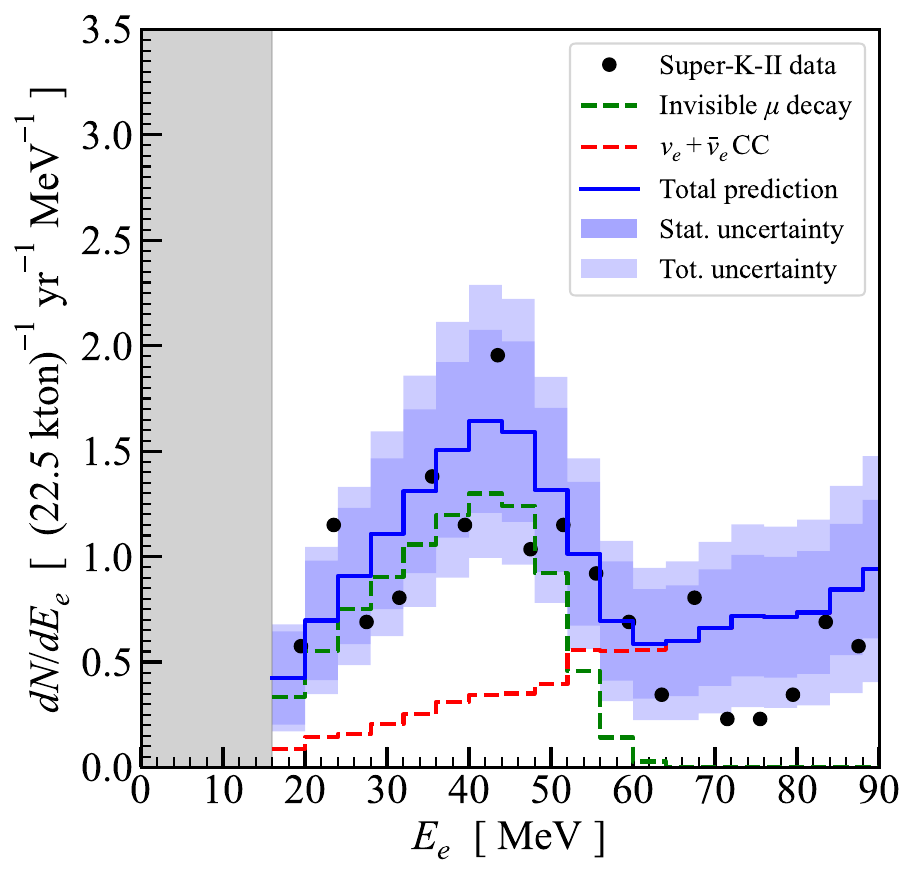}
\includegraphics[width=0.49\textwidth]{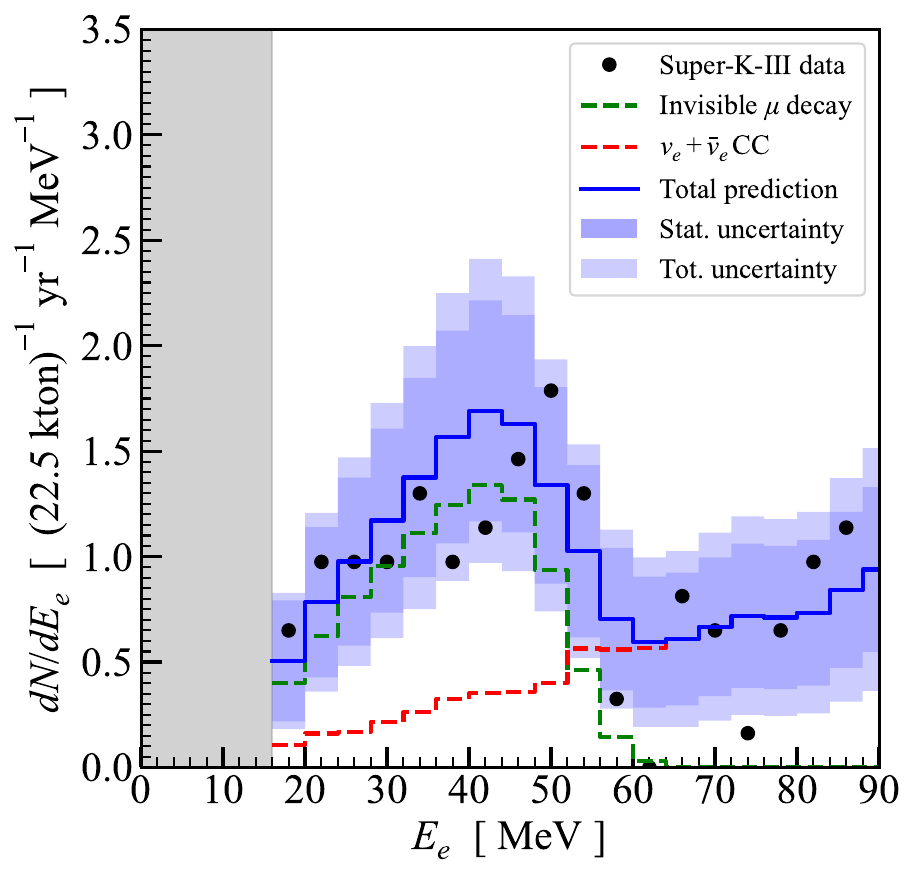}
\includegraphics[width=0.49\textwidth]{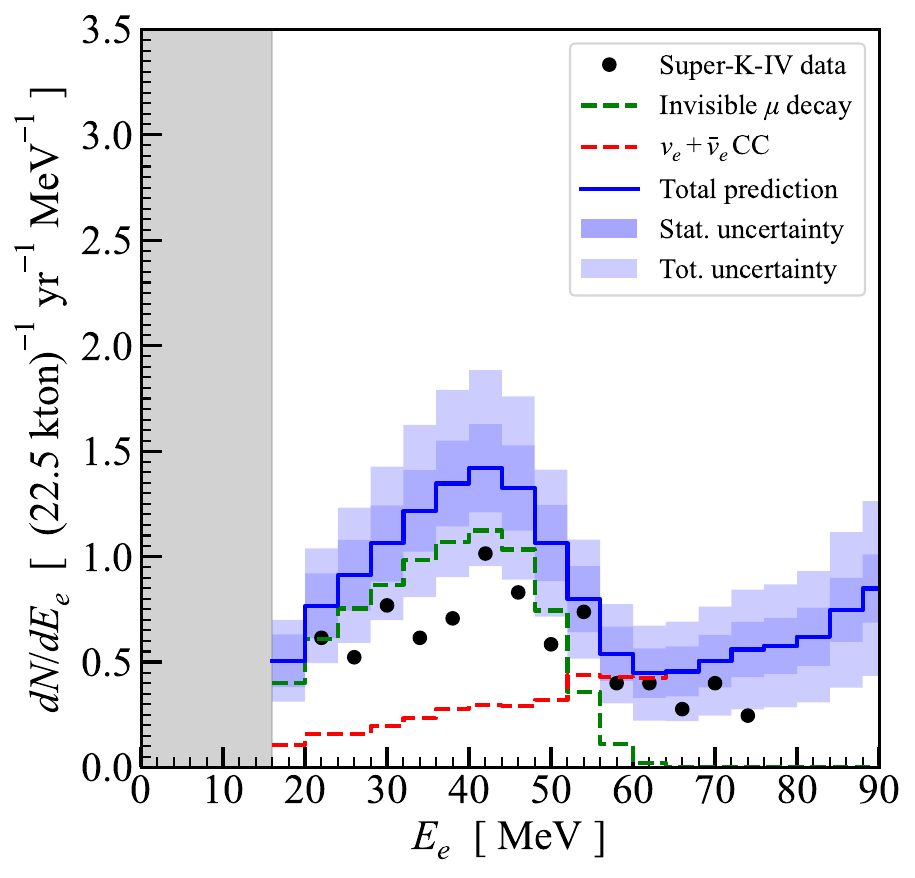}
\caption{Same as Fig.~\ref{fig_reproduce_LE_LFG}, but for the RFG-LS model set. Overall, it gives a higher prediction than the data.}
\label{fig_reproduce_LE_RFG}
\end{figure*}

\begin{figure*}[t]
\includegraphics[width=0.49\textwidth]{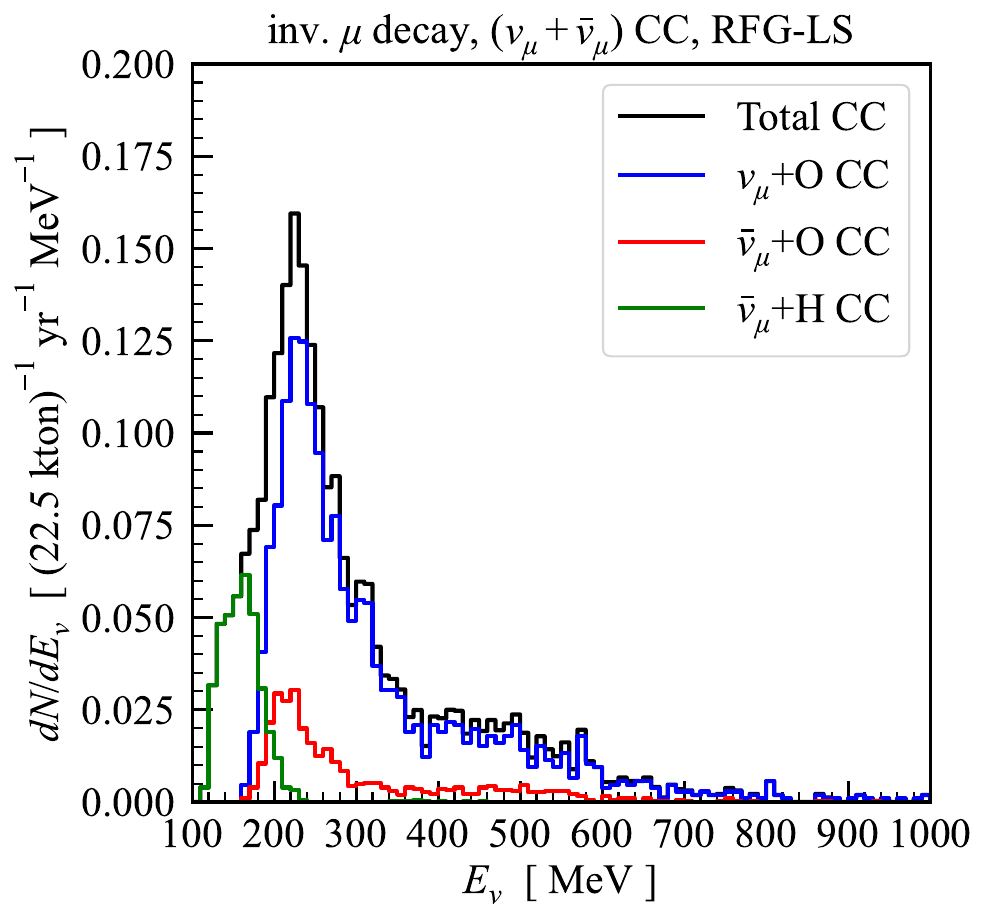}
\includegraphics[width=0.49\textwidth]{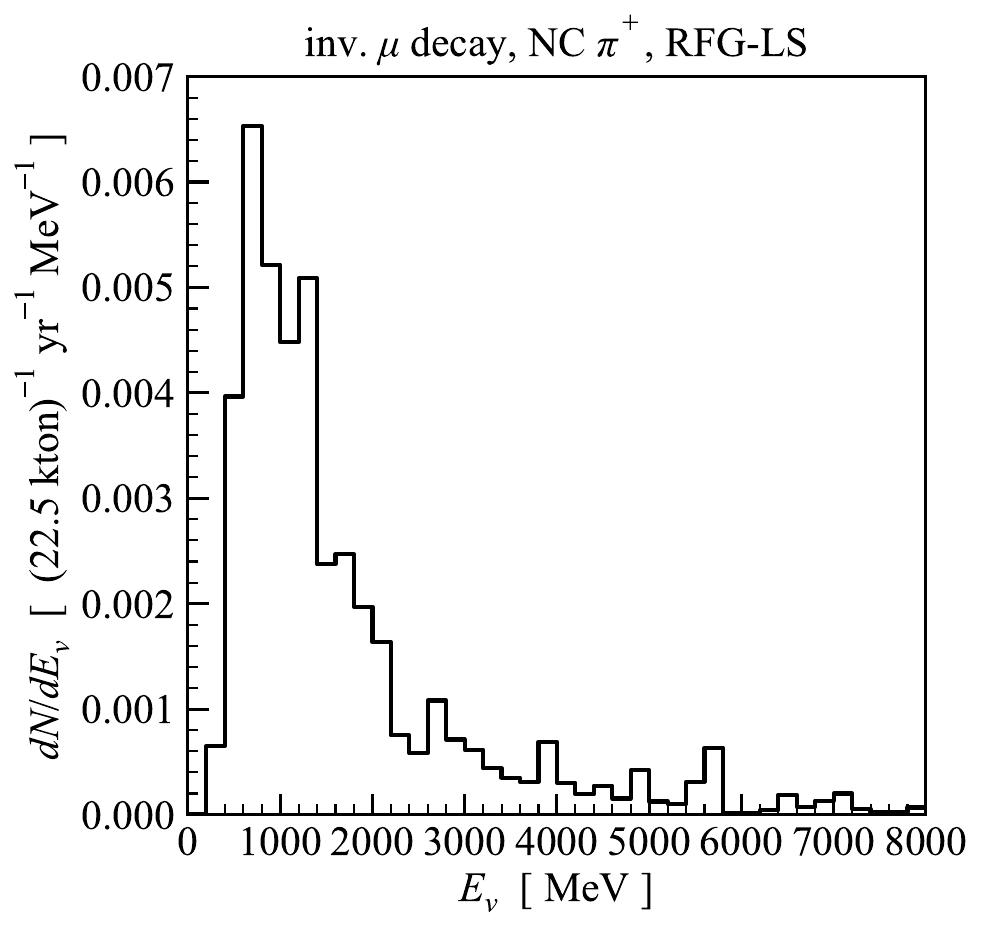}
\caption{ Same as Fig.~\ref{fig_parent_spec_invmu_LFG}, but for the RFG-LS model set.}
\label{fig_parent_spec_invmu_RFG}
\end{figure*}

\begin{figure*}[t]
\includegraphics[width=0.479\textwidth]{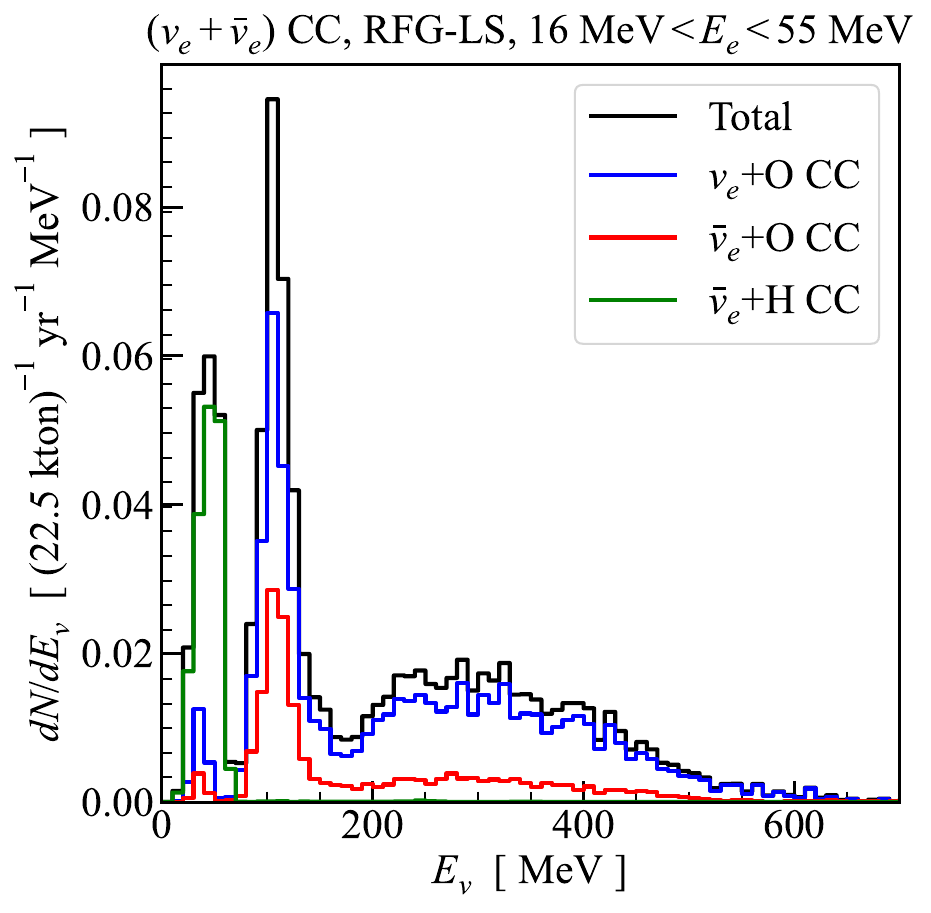}
\includegraphics[width=0.49\textwidth]{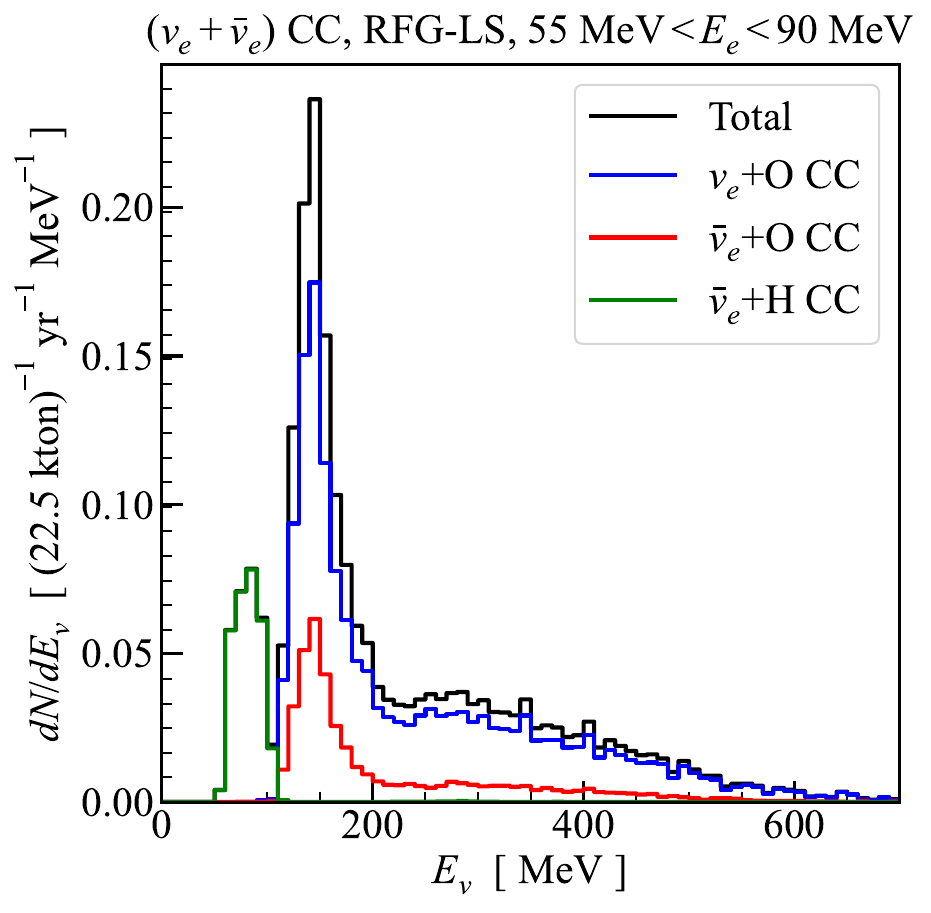}
\caption{Same as Fig.~\ref{fig_parent_spec_nuecc_LFG}, but for the RFG-LS model set.}
\label{fig_parent_spec_nuecc_RFG}
\end{figure*}

\clearpage
\twocolumngrid
\bibliography{references}


\end{document}